\documentclass[12pt]{article}
\usepackage{adjustbox}
\usepackage{bm}
\usepackage{amsfonts}
\usepackage[colorlinks,citecolor=blue,urlcolor=blue,hyperfootnotes=false]{hyperref}
\usepackage{amssymb}
\usepackage[hyperref]{ntheorem}
\usepackage{graphicx}
\usepackage{float}
\usepackage[doublespacing,nodisplayskipstretch ]{setspace}
\usepackage[ a4paper, bottom=0.5in]{geometry}
\usepackage{booktabs}
\usepackage{threeparttable}
\usepackage{amsmath}
\usepackage{lscape}
\usepackage{scrextend}
\usepackage{relsize}
\usepackage{multicol}
\usepackage{chngcntr}
\usepackage{etoolbox}
\usepackage[titletoc, title]{appendix}
\usepackage[sectionbib]{natbib}
\usepackage{threeparttable}
\usepackage{amsmath}
\usepackage{lscape}
\usepackage{relsize}
\usepackage{multicol}
\usepackage{chngcntr}
\usepackage{amssymb}
\usepackage{tabularx}
\usepackage[english]{babel}
\usepackage{multirow}
\usepackage[font=small]{caption}
\usepackage{titlesec}
\usepackage{array}
\usepackage{color}
\usepackage{afterpage}
\usepackage{framed}
\usepackage{scalerel,stackengine}
\usepackage{autobreak}
\usepackage{xr}
\usepackage[sectionbib]{natbib}
\usepackage[sectionbib]{chapterbib}
\usepackage{titlesec}

\allowdisplaybreaks
\newdimen \dummy
\dummy=\oddsidemargin
\titlespacing \section{0pt}{1.0ex plus -1ex minus -.2ex}{-\parskip}
\titlespacing \subsection{0pt}{1.0ex plus -1ex minus -.2ex}{-\parskip}
\AtBeginDocument{
\addtolength{\abovedisplayskip}{-1ex}
\addtolength{\belowdisplayskip}{-1ex}
}
\newcolumntype{x}[1]{>{\centering \arraybackslash \hspace{0pt}}p{#1}}
\theoremstyle{plain}
\newtheorem{theorem}{Theorem}
\newtheorem{corollary}{Corollary}

\newtheorem{lemma}{Lemma}

\newtheorem{assumption}{Assumption}

\theorembodyfont{\normalfont}
\newtheorem{Rem}{Remark}
\newtheorem{algorithm}{Algorithm}
\deffootnote{1em}{1.6em}{\thefootnotemark \enskip}

\newcommand\independent{\protect\mathpalette{\protect\independentT}{\perp}}
\def\independentT#1#2{\mathrel{\rlap{$#1#2$}\mkern2mu{#1#2}}}
\input{tcilatex}
\numberwithin{equation}{section}
\numberwithin{assumption}{section}
\providecommand{\keywords}[1]{\textbf{\textit{Keywords:}} #1}
\providecommand{\U}[1]{\protect\rule{.1in}{.1in}}

\begin{document}

\title{Nonparametric Tests for Treatment Effect Heterogeneity with Duration
Outcomes}
\author{Pedro H. C. Sant'Anna\thanks{
Date: February 8 2020. Contact information: Department of Economics, Vanderbilt
University, VU Station B \#351819, 2301 Vanderbilt Place, Nashville, TN
37235-1819 (e-mail: pedro.h.santanna@vanderbilt.edu). Financial support from
the Spanish Plan Nacional de I+D+I (Grant No. ECO2012-33053 and
ECO2014-55858-P) is acknowledged. This article is a revised version of the
earlier draft entitled \textquotedblleft Nonparametric Tests for Conditional
Treatment Effects with Duration Outcomes\textquotedblright, which is part of
the author's dissertation ``Essays on Duration and Count Data Models,'\
Universidad Carlos II de Madrid in Madrid, Spain, which won the 2016 Zellner
Award, sponsored by the ASA Section on Business and Economic Statistics and
SAS. The author thanks Miguel Delgado, Carlos Velasco, Winfried Stute, Juan
Carlos Escanciano, Tong Li, Atsushi Inoue, and seminar participants at many
seminars/conferences for helpful comments and discussions.} \\
%EndAName
Vanderbilt Univeristy}
\date{}
\maketitle

\begin{onehalfspacing}
\begin{abstract}
This article proposes different tests for treatment
effect heterogeneity when the outcome of interest, typically a duration variable, may be right-censored.
The proposed tests study whether a policy 1)  has zero distributional (average) effect for all
subpopulations defined by covariate values, and 2) has homogeneous average effect across different subpopulations. The proposed
tests are based on two-step Kaplan-Meier integrals and do not rely on
parametric distributional assumptions, shape restrictions, or on
restricting the potential treatment effect heterogeneity across different
subpopulations. Our framework is suitable not only to exogenous
treatment allocation but can also account for treatment noncompliance - an
important feature in many applications. The proposed tests are
consistent against fixed alternatives, and can detect nonparametric
alternatives converging to the null at the parametric $n^{-1/2}$-rate, $n$
being the sample size. Critical values are computed with the assistance of a multiplier bootstrap. The finite sample properties of the proposed tests are examined by means of a Monte Carlo study and
an application about the effect of labor market programs on unemployment duration.
Open-source software is available for implementing all proposed tests.
\\
\keywords{Causal inference; duration data; empirical process; Kaplan-Meier; survival analysis}
\end{abstract}
\end{onehalfspacing}\pagebreak

\section{Introduction}

Assessing whether a policy or treatment has any effect on an outcome of
interest has been one of the main concerns in economics and statistics. As
summarized by \cite{Imbens2009}, the focus of policy evaluation literature
has been mainly confined to identifying and estimating unconditional
treatment effect (TE) measures such as the average, distribution and
quantile treatment effects. However, one important aspect of policy
evaluations is that treatment effects tend to vary across different
subpopulations, and focusing on unconditional\ TE measures can mask
important heterogeneity in policy interventions. For instance, a labor
market program that does not affect the unemployment duration for the
overall population might still be effective for a subgroup of individuals
with specific observable characteristics. Assessing if this is the case is
particularly important for researchers and policymakers interested in
generalizing some findings across time, places and populations, which the
literature calls \textquotedblleft external validity\textquotedblright ; see
e.g. \cite{Hotz2005}, \cite{Bitler2006, Bitler2008, Bitler2014}, \cite%
{Crump2008}, and \cite{Ding2015}. Treatment effect heterogeneity also plays
an important role in designing statistical treatment rules, see e.g. \cite%
{Manski2004}.

In this article, we propose a unified approach to construct tests for
different forms of treatment effect heterogeneity, paying particular
attention to situations in which the outcome of interest, typically a
duration variable, may be subject to right censoring. We develop tests for
both average and distribution treatment effects conditional on covariate
values. In particular, we consider nonparametric tests to assess whether $%
(a) $ there is any particular subpopulation defined by covariates for which
a policy intervention has a nonzero distribution (or average) effect, and $%
(b)$ the average treatment effect vary across different subgroups. All
proposed tests can be applied under unconfounded treatment assignments, see
e.g. \cite{Rosenbaum1983}, but also when selection into treatment is
endogenous and a binary instrumental variable is available to the
researcher, see e.g. \cite{Imbens1994} and \cite{Angrist1996}. Finally, we
emphasize that all proposed tests are also directly applicable to assess TE
heterogeneity with respect to a subset of the available covariate and not
only with respect to the entire vector of observed characteristics; see
Remark \ref{sub}.

The proposed methodology relies on three main components. First, the tests
are based on inverse probability weighted (IPW) estimators, in which the
propensity score is estimated by nonparametric methods. We focus on the
Series Logit Estimator proposed by \cite{Hirano2003}. Second, as we are
interested in TE heterogeneity across subgroups defined by covariates, the
tests are based on conditional moment restrictions. To avoid the use of
smooth estimates and the \textquotedblleft curse of
dimensionality,\textquotedblright\ we adopt an integrated moment approach,
see e.g. \cite{Bierens1982}, \cite{Bierens1997}, \cite{Stute1997}, and \cite%
{Escanciano2006a}. Finally, to tackle the potential censoring problem
inherited in duration outcomes, we characterize the integrated moments as
Kaplan-Meier (KM) integrals, see e.g. \cite{Stute1993a}, \cite{Stute1993,
Stute1995a, Stute1996a}, \cite{Chen1997}, \cite{Sellero2005}, and \cite%
{SantAnna2016}. It is important to emphasize that such an approach is
suitable for both censored and uncensored data.

Combining the aforementioned ingredients, we propose different tests for TE
heterogeneity. Our test statistics are suitable functionals of empirical
processes whose limiting distribution under the null can be estimated using
a multiplicative-type bootstrap. Our proposed tests are of the omnibus type,
i.e., they are consistent against any nonparametric fixed alternative.
Furthermore, they can detect nonparametric local alternatives converging to
the null at the parametric $n^{-1/2}$-rate, $n$ being the sample size. To
the best of our knowledge, no other nonparametric test for TE heterogeneity
share these properties, even when censoring is not an issue.

The closest papers to ours are \cite{Abadie2002}, \cite{Crump2008}, and \cite%
{Lee2009d}. In a context without censoring and covariates, \cite{Abadie2002}
proposes tests for the null hypotheses of zero distribution (local)
treatment effect and first-order stochastic dominance between treatment and
control groups when selection into treatment may be endogenous. Our proposal
generalizes \cite{Abadie2002} by accommodating both covariates (and,
therefore, treatment effect heterogeneity) and randomly censored outcomes. 
\cite{Crump2008} propose smoothed-based tests for the null of hypotheses of
zero and constant conditional average treatment effects under the
unconfoundedness assumption. Our proposal generalizes \cite{Crump2008} by
considering tests for treatment effects heterogeneity beyond the conditional
mean, and by allowing endogenous treatment allocations and censored
outcomes. Finally, \cite{Lee2009d} proposes a Mann--Whitney test for the
null hypothesis of zero conditional distribution treatment effect (like $%
\left( a\right) $ above) for randomly censored outcomes. Nonetheless, it is
not clear how one can generalize the proposal in \cite{Lee2009d} to settings
with endogeneity, or how one can use his approach to test other hypotheses
related to treatment effect heterogeneity like $(b)$. \cite{Delgado2013}, 
\cite{Chang2015}, \cite{Hsu2013}, and \cite{Lee2017} propose alternative
tests for treatment effect heterogeneity but do not allow for right-censored
outcomes.

In summary, we contribute to the literature on different fronts. This paper
is the first to propose a family of nonparametric tests for TE heterogeneity
that $\left( i\right) $ can easily accommodate a variety of research designs
and (random) censoring and $\left( ii\right) $ are able to detect local
alternatives converging to null at the parametric rate. In addition, $\left(
iii\right) $ this paper is one of the first to introduce Kaplan-Meier
integrals to the program evaluation literature; see also \cite{SantAnna2016}
for two-step Kaplan-Meier estimators of different unconditional treatment
effect measures.

The remainder of the paper is organized as follows. To gain intuition, we
first describe the basic setup in which selection into treatment is
exogenous and concentrate on testing the null of zero conditional
distribution treatment effect. In Section 3, we derive the asymptotic
distribution for the baseline tests and introduce a bootstrap method to
approximate their critical values. In Section 4, we present extensions of
our basic setup. We consider the null of zero conditional average treatment
effect and the null of constant average treatment effect across
subpopulations. Furthermore, we show how one can modify the aforementioned
tests to accommodate endogenous treatment allocation. A Monte Carlo study in
Section 5 investigates the finite sample properties of the tests. In Section
6, \ we use data from the Illinois Reemployment Bonus Experiment and apply
the proposed policy evaluation tools to assess the effect of unemployment
insurance bonus on unemployment duration. All mathematical proofs are
gathered in the Appendix. Finally, all tests discussed in this article can
be implemented via the open-source R package \textit{kmte,} which is freely
available from GitHub (\url{https://github.com/pedrohcgs/kmte}).

\textbf{Notation}:\ Let $1\left\{ A\right\} $ be the indicator function,
that is, $1\left\{ A\right\} $ is equal to one if $A$ is true and equal to
zero otherwise. When $A$ is a vector, such function is taken coordinatewise.
For any generic function $J,$ let $J\left( y-\right) =\lim_{a\uparrow
y}J\left( a\right) $, $J\left\{ y\right\} =J\left( y\right) -J\left(
y-\right) $, and denote the continuous part of $J\left( \cdot \right) $ by $%
J^{c}\left( \cdot \right) $. Let $i=\sqrt{-1}$ be the imaginary number.
Denote the support of a generic random variable $Z$ by $\mathcal{X}_{Z}$.
For a set $\mathcal{W}$, let $l^{\infty }\left( \mathcal{W}\right) $ be the
Banach space of all uniformly bounded real functions on $\mathcal{W}$
equipped with the uniform metric $\left\Vert f\right\Vert _{\mathcal{W}%
}\equiv \sup_{z\in \mathcal{W}}\left\vert f\left( z\right) \right\vert $. We
use the notation $\left\Vert \cdot \right\Vert _{\infty }$ to denote the
supremum norm. The symbol $\Rightarrow $ denotes weak convergence in $\left(
l^{\infty }\left( \mathcal{W}\right) ,\mathcal{W}_{\infty }\right) $ in the
sense of J. Hoffmann-J$\phi $rgensen, where $\mathcal{W}_{\infty }$ denotes
the corresponding Borel $\sigma $-algebra, and $\overset{p}{\rightarrow }$
denotes convergence in (outer) probability, see e.g. \cite{VanderVaart1996}.
Throughout the paper, all random variables are defined on a common
probability space $\left( \Omega ,\mathcal{A},\mathbb{P}\right) .$

\section{Testing for zero conditional distribution treatment effect with
censored outcomes\label{Baseline}}

\subsection{Statistical Framework}

We consider a set of individuals flowing into a state of interest, and the
time these individuals spend in that state is the outcome of interest, $Y$.
Upon inflow, an individual participates in the program or not, i.e., he/she
either receives treatment or not. Let $Y\left( 0\right) $ be the potential
outcome if no treatment were received and let $Y\left( 1\right) $ be the
potential outcome if treatment were received. Define $T$ as the treatment
indicator, i.e., $T=1$ if the unit is treated and $T=0$ otherwise. The
realized outcome is $Y=(1-T)Y\left( 0\right) +TY\left( 1\right) $. The
realized outcome, however, is not always observed because of the censoring
mechanism. Let $C\left( 0\right) $ and $C\left( 1\right) $ be potential
censoring random variables under the control and treatment groups,
respectively, and $C=(1-T)C\left( 0\right) +TC\left( 1\right) $ be the
realized censored variable, beyond which $Y$ is not observed. For example, $%
C $ may be the time from treatment assignment until the end of a follow-up.
The observed outcome is $Q=(1-T)Q\left( 0\right) +TQ\left( 1\right) $ where $%
Q\left( t\right) =\min \left( Y\left( t\right) ,C\left( t\right) \right) $, $%
t\in \left\{ 0,1\right\} $. On top of $Q$, the non-censoring indicator $%
\delta =(1-T)\delta \left( 0\right) +T\delta \left( 1\right) ,$ $\delta
\left( t\right) =1\left\{ Y\left( t\right) \leq C\left( t\right) \right\}
,~t\in \left\{ 0,1\right\} $, and a vector of pre-treatment variables $%
\mathbf{X}$ are also observed. We consider $\left\{ \left( Q_{i},\delta
_{i},T_{i},\mathbf{X}_{i}\right) \right\} _{i=1}^{n}$ as independent and
identically distributed ($iid$) random variables.

Denote the conditional distribution of potential outcomes $Y\left( 0\right) $
and $Y\left( 1\right) $ by $F_{Y\left( 0\right) |\mathbf{X}}\left( y|\mathbf{%
\cdot }\right) $ and $F_{Y\left( 1\right) |\mathbf{X}}\left( y|\mathbf{\cdot 
}\right) $, respectively, and let the conditional distribution treatment
effect be defined as $\Upsilon \left( y|\cdot \right) \equiv F_{Y\left(
1\right) |\mathbf{X}}\left( y|\cdot \right) -F_{Y\left( 0\right) |\mathbf{X}%
}\left( y|\cdot \right) .$ To gain intuition, we first focus on testing the
hypothesis that the distribution treatment effect (DTE) is equal to zero for
every subpopulation defined by covariates, that is,%
\begin{equation}
H_{0}:\Upsilon \left( y|\mathbf{X}\right) =0\text{ }a.s.\text{ }\forall y\in 
\mathcal{W}_{Y},  \label{cste}
\end{equation}%
where $\mathcal{W}_{Y}$ $\subset $ $\mathcal{X}_{Y}$. The alternative
hypothesis $H_{1}$ is the negation of $H_{0}$.

A crucial step towards testing (\ref{cste}) is to show that $\Upsilon \left(
y|\mathbf{\cdot }\right) $ can be identified from the data. To this end, we
make the following assumptions.

\begin{assumption}
\label{unconfound}$\left( i\right) $ $\left( Y\left( 0\right) ,Y\left(
1\right) ,C\left( 0\right) ,C\left( 1\right) \right) 
%TCIMACRO{\TeXButton{indep}{\independent}}%
%BeginExpansion
\independent%
%EndExpansion
T|\mathbf{X}$~$a.s\mathbf{.;}$ and $\left( ii\right) $ for some $\varepsilon
>0$, $\varepsilon \leq \mathbb{P}\left( T=1|\mathbf{X}\right) \leq
1-\varepsilon ~a.s..;$
\end{assumption}

\begin{assumption}
\label{censoring_identification} $(i)$ $\left( Y\left( 0\right) ,Y\left(
1\right) \right) $ $%
%TCIMACRO{\TeXButton{indep}{\independent}}%
%BeginExpansion
\independent%
%EndExpansion
\left( C\left( 0\right) ,C\left( 1\right) \right) |T;$ and $\left( ii\right) 
$ for $t\in \left\{ 0,1\right\} $,\linebreak\ $\mathbb{P}\left( Y\left(
t\right) \leq C\left( t\right) |\mathbf{X},T,Y\left( t\right) \right) =%
\mathbb{P}\left( Y\left( t\right) \leq C\left( t\right) |T,Y\left( t\right)
\right) $ $a.s..$
\end{assumption}

We will use the shortcut notation $p_{0}\left( \mathbf{x}\right) \equiv 
\mathbb{P}(T=1|\mathbf{X=x})$ and refer to $p_{0}\left( \mathbf{x}\right) $
as the (true) propensity score. Assumption \ref{unconfound} is standard in
the treatment effects literature. Assumption \ref{unconfound}$\left(
i\right) $ states that, conditional on observables, treatment assignment is
independent of potential outcomes and censoring. Assumption \ref{unconfound}$%
\left( ii\right) $ states that there is overlap in the covariate
distributions.

In the absence of censoring, \cite{Rosenbaum1983} show that Assumption \ref%
{unconfound} suffices to identify different treatment effect measures, in
particular $\Upsilon \left( y|\mathbf{\cdot }\right) $. Nonetheless, in our
setup censoring introduces an additional identification challenge because
the probability of being censored is related to potential outcomes, that is,
censoring occurs only if $Y\left( t\right) >C\left( t\right) ,$ $t\in
\left\{ 0,1\right\} $. Ignoring the censoring problem or analyzing only the
uncensored outcomes would, therefore, introduce another source of
confounding. To circumvent this problem, Assumption \ref%
{censoring_identification} imposes additional structure on the censoring
mechanism.

Assumption \ref{censoring_identification} states that, conditionally on the
treatment status, the potential outcomes are independent of the potential
censoring random variables, and that, given the underlying potential
outcome\ $Y\left( t\right) $, $t\in \left\{ 0,1\right\} $, and the treatment
status $T$, the covariates do not provide any further information on whether
censoring will take place. A particular case in which Assumption \ref%
{censoring_identification} is satisfied is when $C$ is independent of $%
\left( Y,\mathbf{X},T\right) $, as assumed by \cite{Bang2000}, \cite%
{Anstrom2001}, \cite{Honore2002}, \cite{Lee2005}, \cite{Blundell2007}, among
many others. One must bear in mind that Assumption \ref%
{censoring_identification} is more general than this particular case: it
does not impose any restriction on how $\left( Y\left( 1\right) ,Y\left(
0\right) \right) $ and $\left( C\left( 1\right) ,C\left( 0\right) \right) $
depend on $T$, it allows some dependence between $C\left( 1\right) $, $%
C\left( 0\right) $, $T$ and $\mathbf{X}$, and allows the occurrence of
censoring to depend on $\mathbf{X}$.

In the following, we establish that, given Assumptions \ref{unconfound}-\ref%
{censoring_identification}, a variety of TE measures are identified from $%
\left( Q,\delta ,T,\mathbf{X}\right) .$ In particular, we show that the
joint distribution of potential outcome $Y\left( t\right) $ and the vector
of covariates $\mathbf{X}$, denoted by $F_{Y\left( t\right) ,\mathbf{X}%
}\left( y,\mathbf{x}\right) =\mathbb{P}(Y\left( t\right) \leq y,\mathbf{X}%
\leq \mathbf{x})$, $t\in \left\{ 0,1\right\} $ is identified. Once $%
F_{Y\left( t\right) ,\mathbf{X}}\left( y,\mathbf{x}\right) $ is identified, $%
F_{Y\left( t\right) |\mathbf{X}}\left( y|\mathbf{\cdot }\right) $ can be
recovered by taking the appropriate Radon-Nikodym derivative.

We\ now introduce the multivariate Kaplan-Meier joint distribution, which is
the key piece to characterize our Kaplan-Meier integrals. Toward this end,
let $H_{Q,\mathbf{X}|T}(y,\mathbf{x}|t)=\mathbb{P}(Q\leq y,\mathbf{X}\leq 
\mathbf{x}|T=t)$, $H_{Q,\mathbf{X}|T}^{1}(y,\mathbf{x}|t)=\mathbb{P}(Q\leq y,%
\mathbf{X}\leq \mathbf{x},\delta =1|T=t)$ and 
\begin{equation}
\Lambda \left( y,\mathbf{x}|t\right) =\int_{0}^{y}\frac{H_{Q,\mathbf{X}%
|T}^{1}(d\bar{y},\mathbf{x}|t)}{1-H_{Q,\mathbf{X}|T}(\bar{y}-,\boldsymbol{%
\infty }|t)}.  \label{cumhazard}
\end{equation}%
For $t\in \left\{ 0,1\right\} $, let $\tau _{\left( t\right) }:=\min \left(
\tau _{Y\left( t\right) },\tau _{C\left( t\right) }\right) $, where $\tau
_{Y\left( t\right) }=\inf $ $\left\{ y:\mathbb{P}\left( Y\left( t\right)
\leq y\right) =1\right\} ,$ and $\tau _{C\left( t\right) }=\inf $ $\left\{ y:%
\mathbb{P}\left( C\left( t\right) \leq y\right) =1\right\} $ are the least
upper bound of the support of $Y\left( t\right) $ and $C\left( t\right) $.
For simplicity, assume that $\tau _{C\left( 1\right) }=\tau _{C\left(
0\right) }=\tau _{C}$, $\tau _{Y\left( 1\right) }=\tau _{Y\left( 0\right)
}=\tau _{Y}$, implying that $\tau _{\left( 1\right) }=\tau _{\left( 0\right)
}=\tau $.\ For $t\in \left\{ 0,1\right\} $, denote by $A\left( t\right) $
the (possibly empty) set of atoms of the cumulative distribution function of 
$Q\left( t\right) $, and let 
\begin{equation}
F_{Q,\mathbf{X}|T}^{km}\left( y,\mathbf{x}|t\right) =1-\exp \left( -\Lambda
^{c}\left( y,\mathbf{x}|t\right) \right) \dprod_{\bar{y}\leq y}\left[
1-\Lambda \left( \left\{ \bar{y}\right\} ,\mathbf{x}|t\right) \right] ,
\label{kmproduct}
\end{equation}%
where $\Lambda \left( \left\{ \bar{y}\right\} ,\mathbf{x}|t\right) \equiv
\Lambda \left( \bar{y},\mathbf{x}|t\right) -\Lambda \left( \bar{y}-,\mathbf{x%
}|t\right) $ denotes the jump size (mass) of $\Lambda \left( y,\mathbf{x}%
|t\right) $ at $\bar{y}$. In the case where $Q$ is absolutely continuous, $%
\Lambda \left( \left\{ \bar{y}\right\} ,\mathbf{x}|t\right) \equiv 0$ for
almost all $\bar{y},\mathbf{x}$ and $t$. In the other extreme case where $Q$
is purely discrete with support points $\left\{ y_{1},y_{2},\dots
,y_{s}\right\} $, $\Lambda ^{c}\left( y,\mathbf{x}|t\right) \equiv 0~$for
almost all $y,\mathbf{x}$ and $t$, and 
\begin{equation*}
\prod_{\bar{y}\leq y}\left[ 1-\Lambda \left( \left\{ \bar{y}\right\} ,%
\mathbf{x}|t\right) \right] \equiv \left[ 1-\Lambda \left( \left\{
y_{1}\right\} ,\mathbf{x}|t\right) \right] ^{1\left\{ y_{1}\leq y\right\}
}\times \dots \times \left[ 1-\Lambda \left( \left\{ y_{s}\right\} ,\mathbf{x%
}|t\right) \right] ^{1\left\{ y_{s}\leq y\right\} }.
\end{equation*}%
Note that (\ref{kmproduct}) allows $Q$ to have both discrete and absolutely
continuous components.

Finally, for any measurable function $g\left( \cdot \right) ,${%
\begin{multline}
\mathbb{E}^{km}\left[ g\left( Q,\mathbf{X},T\right) \right] \equiv \int
g\left( T,\mathbf{X},1\right) F_{Q,\mathbf{X}|T}^{km}\left( d\bar{y},d%
\mathbf{\bar{x}}|1\right) \mathbb{P}\left( T=1\right)  \label{kmint} \\
+\int g\left( Q,\mathbf{X},0\right) F_{Q,\mathbf{X}|T}^{km}\left( d\bar{y},d%
\mathbf{\bar{x}}|0\right) \mathbb{P}\left( T=0\right) .
\end{multline}%
}

\begin{lemma}
\label{identification}Suppose Assumptions \ref{unconfound}-\ref%
{censoring_identification} hold. Let $h\left( \cdot \right) $ be any
measurable function of $\left( Y,\mathbf{X},T\right) $ such that $\mathbb{E}%
\left[ \left\vert h\left( Y,\mathbf{X},T\right) \right\vert \right] <\infty $%
. Then, for $t\in \left\{ 0,1\right\} $,%
\begin{multline}
\mathbb{E}^{km}\left[ \dfrac{1\left\{ T=t\right\} h\left( Q,\mathbf{X,}%
T\right) }{\mathbb{P}\left( T=t|\mathbf{X}\right) }\right] =\mathbb{E}\left[
h\left( Y\left( t\right) ,\mathbf{X,}t\right) 1\left\{ Y\left( t\right)
<\tau \right\} \right] +  \label{ident.functions} \\
1\left\{ \tau \in A\left( t\right) \right\} \mathbb{E}\left[ 1\left\{
Y\left( t\right) =\tau \right\} h\left( \tau ,\mathbf{X,}t\right) \right] .
\end{multline}%
Moreover, (\ref{ident.functions}) also holds conditional on $\mathbf{X}$.
\end{lemma}

Lemma \ref{identification} is based on \cite{SantAnna2016}, and extends \cite%
{Rosenbaum1983} identification results to setups with censored outcomes. It
relies on replacing $F_{Y,\mathbf{X}|T}$ by the multivariate Kaplan-Meier $%
F_{Q,\mathbf{X}|T}^{km}$ in (\ref{kmint}). Note that $F_{Q,\mathbf{X}%
|T}^{km} $ only depends on $\left( Q,\delta ,T,\mathbf{X}\right) $ and,
therefore, is self-adjusted to the censoring problem. Furthermore, in the
absence of censoring, $F_{Q,\mathbf{X}|T}^{km}=F_{Y,\mathbf{X}|T}~a.s.$ \cite%
[Proposition 1, pg. 301]{Shorack1986}, implying that in such a case, our
identification result reduces to those of \cite{Rosenbaum1983}. On the other
hand, when the outcome of interest is subject to random censoring but one
ignores this feature and treats $Q$ as $Y$, the \textquotedblleft
standard\textquotedblright\ IPW estimand can be expressed as 
\begin{multline*}
\mathbb{E}\left[ \frac{1\left\{ T=t\right\} h\left( Q,X,T\right) }{\mathbb{P}%
\left( T=t|X\right) }\right] =\mathbb{E}\left[ h\left( Y\left( t\right)
,X,t\right) |Y\left( t\right) \leq C\left( t\right) \right] \mathbb{P}\left(
Y\left( t\right) \leq C\left( t\right) \right) \\
+\mathbb{E}\left[ h\left( C\left( t\right) ,X,t\right) |Y\left( t\right)
>C\left( t\right) \right] \mathbb{P}\left( Y\left( t\right) >C\left(
t\right) \right) .
\end{multline*}%
In other words, if one ignores the censoring problem, one would only
identify a weighted average of functionals of $\left( i\right) $ $\left(
Y\left( t\right) ,X,t\right) $ for the uncensored subpopulation ($Y\left(
t\right) \leq C\left( t\right) ~a.s.$) and $\left( ii\right) $ $\left(
C\left( t\right) ,X,t\right) $ for the censored subpopulation ($Y\left(
t\right) >C\left( t\right) ~a.s.$). Such weighted average representations do
not have any clear causal interpretation as these subpopulations (being
censored or not) are directly related to the potential outcomes. We bypass
the additional source of confounding introduced by the censoring problems by
using the multivariate Kaplan-Meier distribution $F_{Q,\mathbf{X}|T}^{km}$
as an integrating measure.

It is important to remark that, because of the censoring problem,
nonparametric identification of statistical characteristics that depend on
the entire support of $Y\left( t\right) $ such as $\mathbb{E}\left[ Y\left(
t\right) \right] $, $t\in \left\{ 0,1\right\} $, crucially depends on the
local structure of the distribution of $Y\left( t\right) $ and $C\left(
t\right) $ at their endpoint. If $\tau _{Y}<\tau _{C}$, identification is
guaranteed for \textit{any (}well defined) functional of interest; if $\tau
_{Y}>\tau _{C}$, only restricted moments can be identified; and if $\tau
_{C}=\tau _{Y}$, identification is guaranteed unless 
\begin{equation}
F_{Y\left( t\right) ,\mathbf{X}}\left( \left\{ \tau \right\} ,\mathbf{x}%
\right) >0~and~\mathbb{P}\left( C\left( t\right) <\tau \right) =1,
\label{case1}
\end{equation}%
where $F_{Y\left( t\right) ,\mathbf{X}}\left( \left\{ \tau \right\} ,\mathbf{%
x}\right) \equiv F_{Y\left( t\right) ,\mathbf{X}}\left( \tau ,\mathbf{x}%
\right) -F_{Y\left( t\right) ,\mathbf{X}}\left( \tau -,\mathbf{x}\right) .$
In particular, whenever $Y\left( t\right) $ is continuous, $\mathbb{E}\left[
h\left( Y\left( t\right) ,\mathbf{X,}t\right) \right] $ is nonparametrically
identified if $\tau _{Y}\leq \tau _{C}$, otherwise one can only identify $%
\mathbb{E}\left[ h\left( Y\left( t\right) ,\mathbf{X,}t\right) 1\left\{
Y\left( t\right) \leq \tau \right\} \right] .$ This is intuitive because
relevant information about $F_{Y\left( t\right) ,\mathbf{X}}$ on $(\tau
_{C},\tau _{Y}]$ will always be cut off as a result of the censoring. Such
information cannot be recovered unless one is willing to rely on additional
parametric/shape restrictions. As indicated by (\ref{ident.functions}) and (%
\ref{case1}), the case of discrete potential outcomes and potential
censoring random variables is more delicate: if the the underlying
distribution of $Y\left( t\right) $ has a positive mass point at $\tau $ but
the $\mathbb{P}\left( C\left( t\right) \geq \tau \right) =0$, no information
about $Y\left( t\right) $ at $\tau $ can be recovered from $\left( Q,\mathbf{%
X},T\right) $\footnote{%
Under Assumptions \ref{unconfound} and \ref{censoring_identification}, and
with no covariates, it is easy to see the lack of identification at the
boundary of the support may appear when $H$ is not absolutely continuous. In
light of (\ref{cumhazard}) and (\ref{kmproduct}), if $F_{Y\left( t\right) }$
has a mass point at $\tau $, $F_{Y\left( t\right) }\left( \left\{ \tau
\right\} \right) =F_{Y\left( t\right) }\left( \tau \right) -F_{Y\left(
t\right) }\left( \tau -\right) $, one would naturally hope to identify if by
using integrals of $H_{t,1}\left( \tau \right) -H_{t,1}\left( \tau -\right)
=-\left( \left( 1-H_{t,1}\left( \tau \right) \right) -1-H_{t,1}\left( \tau
-\right) \right) ,$ $t=0,1$. From the definition of $\tau $, we have that,
for $t=0,1$, $0=1-H_{Q|T}\left( \tau |t\right) =\left( 1-F_{Y\left( t\right)
}\left( \tau \right) \right) \left( 1-G_{C\left( t\right) }\left( \tau
\right) \right) $. When $F_{Y\left( t\right) }\left( \tau -\right) <1$ but $%
G_{C\left( t\right) }\left( \tau -\right) =1$, we have that $\tau _{\left(
t\right) }\not\in A\left( t\right) $, and $1-H_{Q|T}\left( \tau -|t\right)
=\left( 1-F_{Y\left( t\right) }\left( \tau -\right) \right) \times 0=0$.
Thus, $H_{t,1}\left( \tau \right) -H_{t,1}\left( \tau -\right) =0,$ implying
that if $Y\left( t\right) $ has a mass point at $\tau \not\in A\left(
t\right) $, one cannot recover information about it from $\left( Q,T\right) $%
. On the other hand, if $F_{Y\left( t\right) }\left( \tau -\right) <1$ but $%
G_{C\left( t\right) }\left( \tau -\right) <1,$ then the identification
follows naturally as one would be able to observe $Q=\tau $ for those with $%
\delta =1$.}. As a direct consequence, in such a case, only restricted
moments of potential outcomes can be nonparametrically point identified.
Indeed, the second term in (\ref{ident.functions}) emphasizes this feature.
In order to ease notation and avoid repetition of arguments, in the rest of
the paper, we rule out (\ref{case1}).

One should bear in mind that although nonparametric identification of
general statistical characteristics is not always guaranteed, Lemma \ref%
{identification} is still very powerful. For instance, applying Lemma \ref%
{identification} with $h\left( Y,\mathbf{X,}T\right) =$ $1\left\{ Y\leq
y\right\} 1\left\{ \mathbf{X}\leq \mathbf{x}\right\} ,$ we get that $%
F_{Y\left( t\right) ,\mathbf{X}}\left( y,\mathbf{x}\right) $ and $F_{Y\left(
t\right) |\mathbf{X}}\left( y|\mathbf{x}\right) $ are identified for $\left(
y,\mathbf{x}\right) \in \lbrack -\infty ,\tau ]\times \mathcal{X}_{_{X}}$.
This is in contrast to the results of \cite{Frangakis1999}, \cite%
{Anstrom2001}, and \cite{Frandsen2014}, who restrict the analysis to $y\in
\lbrack -\infty ,\bar{\tau}]$, with $\bar{\tau}<\tau $. In practice, given
that there is no general rule on how to appropriately choose $\bar{\tau}$,
an ad hoc choice of \textquotedblleft small\textquotedblright\ $\bar{\tau}$
can lead to undesirable loss of information. The results in Lemma \ref%
{identification} completely avoid such a drawback\footnote{%
The aforementioned papers rely on inverse probability weighting techniques
to achieve identification. As such, when $y$ approaches $\tau ,~$the
probability of censoring approaches one, implying that the associated
inverse probability of censoring weights approach infinity. Therefore, it is
not straightforward to see how one can adapt the identification arguments
used in these papers to the case where $\tau =\infty $, for example.
Finally, we note that these boundary problems can become even more perverse
when one turns attention to estimation.}.

\subsection{Characterization of the null hypothesis}

From Lemma \ref{identification}, we have that, for $y\in \lbrack -\infty
,\tau ]$, the conditional DTE ${\small \Upsilon }\left( y|\cdot \right) $ is
identified from the data and, therefore, we are able to characterize the
null hypothesis (\ref{cste}) in terms of observables.

One approach to constructing tests for (\ref{cste}) is to combine Lemma \ref%
{identification} with smoothing techniques, estimate the conditional DTE $%
\Upsilon \left( y|\mathbf{\cdot }\right) $, and then compare how close $%
\Upsilon \left( y|\mathbf{\cdot }\right) $ is to zero. The main drawback of
this strategy\ is that, when the dimension of covariates $\mathbf{X}$ is
moderate as is commonly the case in policy evaluation, tests based on this
local approach suffers from the \textquotedblleft curse of
dimensionality\textquotedblright ; see e.g. \cite{Fan1996} for related tests
in a different context. In the next Lemma, we show that, by exploiting
alternative characterizations of (\ref{cste}), one can avoid estimating $%
\Upsilon \left( y|\mathbf{\cdot }\right) $, alleviating the drawback
associated with the local approach described above. To do so, we rely on the
\textquotedblleft integrated moment approach\textquotedblright\ used in the
goodness-of-fit test literature, see e.g. \cite{Bierens1982}, \cite%
{Stute1997}, \cite{Escanciano2006a}, among others.

\begin{lemma}
\label{integrated}Suppose Assumptions \ref{unconfound}-\ref%
{censoring_identification} hold. Assume that the parametric family \newline
$\mathcal{F}=\left\{ w\left( \mathbf{X},\mathbf{x}\right) :\mathbf{x}\in \Pi
\subset \left[ -\infty ,\infty \right] ^{k}\right\} $ satisfy Assumption \ref%
{w-family} stated in Appendix \ref{App-Assumption}. Then,%
\begin{equation}
\Upsilon \left( y|\mathbf{X}\right) =0\text{ }a.s.\text{ }\forall y\in
\lbrack -\infty ,\tau ]\Leftrightarrow I_{w}\left( y,\mathbf{x}\right) =0%
\text{ }a.e.\text{ in }[-\infty ,\tau ]\times \Pi  \label{equ1}
\end{equation}%
where $\Pi $ is a properly chosen space that may depend on the choice of $w$%
, $I_{w}\left( y,\mathbf{x}\right) =I_{w}^{1}\left( y,\mathbf{x}\right)
-I_{w}^{0}\left( y,\mathbf{x}\right) $, and for $t\in \left\{ 0,1\right\} ,$ 
\begin{equation*}
I_{w}^{t}\left( y,\mathbf{x}\right) \equiv \mathbb{E}^{km}\left[ \dfrac{%
1\left\{ T=t\right\} 1\left\{ Q\leq y\right\} }{\mathbb{P}\left( T=t|\mathbf{%
X}\right) }w\left( \mathbf{X},\mathbf{x}\right) \right] ,
\end{equation*}
\end{lemma}

Lemma \ref{integrated} adapts Lemma 1 of \cite{Escanciano2006a} to the
present context. Examples of parametric families $w\left( \cdot ,\mathbf{x}%
\right) $ such that the equivalence (\ref{equ1}) holds are the exponential
function $w\left( \mathbf{X},\mathbf{x}\right) =\exp (i\mathbf{x}^{\prime }%
\mathbf{X})$ with $\mathbf{x}\in \mathbb{R}^{k}$, as in \cite{Bierens1982},
and the indicator function $w\left( \mathbf{X},\mathbf{x}\right) =1\left\{ 
\mathbf{X}\leq \mathbf{x}\right\} $ with $\mathbf{x}\in \mathcal{X}_{X},$ as
in \cite{Stute1997}; for alternative weight functions, see e.g. \cite%
{Stinchcombe1998}.

From (\ref{equ1}), it is indeed clear that in order to test for (\ref{cste}%
), it suffices to estimate $I_{w}\left( y,\mathbf{x}\right) $ and check if
it is \textquotedblleft sufficiently close to zero\textquotedblright\ for
all values of $\left( y,\mathbf{x}\right) \in \mathcal{W}\subseteq [-\infty
,\tau ]\times \Pi $. Of course, the notion of \textquotedblleft sufficiently
close\textquotedblright\ depends on $(i)$ the norm one uses, say, the $L_{2}$%
-norm, which leads to a Cram\'{e}r-von Mises-type test, or the $\sup $-norm,
which leads to a Kolmogorov-Smirnov-type test; and $\left( ii\right) $ the
sampling distribution of the estimator of $I_{w}\left( y,\mathbf{x}\right) $%
. In the next section, we discuss in detail how one can actually estimate $%
I_{w}\left( y,\mathbf{x}\right) $, and how one can use these estimators to
form test-statistics for (\ref{cste}).

\subsection{Test statistic\label{teststat}}

The characterization of the null hypothesis in (\ref{equ1}) suggests using
functionals of an estimator of $I_{w}\left( \cdot ,\cdot \right) $ as test
statistics. Therefore, we must first estimate $I_{w}\left( \cdot ,\cdot
\right) $ using the sample $\left\{ \left( Q_{i},\delta _{i},T_{i},\mathbf{X}%
_{i}\right) \right\} _{i=1}^{n}$. From Lemma \ref{identification} and the
Total Law of Probability, we have that, for $\left( y,\mathbf{x}\right) \in
\lbrack -\infty ,\tau ]\times \Pi \mathcal{,}\ t\in \left\{ 0,1\right\} ,$ 
\begin{equation}
I_{w}^{t}\left( y,\mathbf{x}\right) =\mathbb{P}\left( T=t\right) \int \dfrac{%
1\left\{ \bar{y}\leq y\right\} w\left( \mathbf{\bar{x}},\mathbf{x}\right) }{%
\mathbb{P}\left( T=t|\mathbf{X}=\mathbf{\bar{x}}\right) }F_{Q,\mathbf{X}%
|T}^{km}\left( d\bar{y},d\mathbf{\bar{x}}|t\right) .  \label{Ieq}
\end{equation}%
Thus, to estimate $I_{w}\left( \cdot ,\cdot \right) $, we have to estimate $%
\mathbb{P}\left( T=t|\mathbf{X}\right) ,$ $F_{Q,\mathbf{X}|T}^{km}\left( y,%
\mathbf{x}|t\right) $ and $\mathbb{P}\left( T=t\right) $, $t\in \left\{
0,1\right\} $.

The task of estimating the propensity score $p_{0}\left( \cdot \right) $ is
relatively standard. For instance, when the data comes from a randomized
experiment, $p_{0}\left( \cdot \right) $ can be estimated by $%
n^{-1}\sum_{i=1}^{n}T_{i}$. Alternatively, when the treatment allocation
depends on observable characteristics, one can nonparametrically estimate $%
p_{0}\left( \cdot \right) $ using the Series Logit Estimator (SLE) proposed
by \cite{Hirano2003}. To define the SLE, let $\boldsymbol{\lambda }=\left(
\lambda _{1},\dots ,\lambda _{r}\right) ^{\prime }$ be a $r$-dimensional
vector of non-negative integers with norm $\left\vert \boldsymbol{\lambda }%
\right\vert =\sum_{j=1}^{r}\lambda _{j}$. Let $\left\{ \boldsymbol{\lambda }%
\left( l\right) \right\} _{l=1}^{\infty }$ be a sequence including all
distinct multi-indices $\boldsymbol{\lambda }$ such that $\left\vert 
\boldsymbol{\lambda }\left( l\right) \right\vert $ is non-decreasing in $l$
and let $\mathbf{x}^{\lambda }=\prod_{j=1}^{r}\mathbf{x}_{j}^{\lambda _{j}}$%
. For any integer $L$, define $\mathbf{R}^{L}\left( \mathbf{x}\right)
=\left( \mathbf{x}^{\lambda \left( 1\right) }\mathbf{,}\dots ,\mathbf{x}%
^{\lambda \left( L\right) }\right) ^{\prime }$ as a vector of power
functions. Let $\mathcal{L}\left( a\right) =\exp \left( a\right) /\left(
1+\exp \left( a\right) \right) $ be the logistic $CDF$. The SLE for $%
p_{0}\left( \mathbf{x}\right) $ is defined as $\hat{p}_{n}\left( \mathbf{x}%
\right) =\mathcal{L}\left( \mathbf{R}^{L}\left( x\right) ^{\prime }%
\boldsymbol{\hat{\pi}}_{L}\right) $, where 
\begin{equation}
\boldsymbol{\hat{\pi}}_{L}=\arg \max_{\boldsymbol{\pi }_{L}}\frac{1}{n}%
\sum_{i=1}^{n}T_{i}\log \left( \mathcal{L}\left( \mathbf{R}^{L}\left( 
\boldsymbol{X}_{i}\right) ^{\prime }\boldsymbol{\pi }_{L}\right) \right)
+\left( 1-T_{i}\right) \log \left( 1-\mathcal{L}\left( \mathbf{R}^{L}\left( 
\boldsymbol{X}_{i}\right) ^{\prime }\boldsymbol{\pi }_{L}\right) \right) .
\label{pscore-series}
\end{equation}%
We write $\mathbb{\hat{P}}_{n}\left( T=1|\mathbf{X}=\mathbf{x}\right) =\hat{p%
}_{n}\left( \mathbf{x}\right) $ and $\mathbb{\hat{P}}_{n}\left( T=0|\mathbf{X%
}=\mathbf{x}\right) =1-\hat{p}_{n}\left( \mathbf{x}\right) $.

Next, we move to the most challenging step: estimating $F_{Q,\mathbf{X}%
|T}^{km}\left( y,\mathbf{x}|t\right) $. Note that, as a result of the binary
nature of $T$, we have to estimate two distribution functions: $F_{Q,\mathbf{%
X}|T}^{km}\left( y,\mathbf{x}|1\right) $, and $F_{Q,\mathbf{X}|T}^{km}\left(
y,\mathbf{x}|0\right) $. To this end, we divide the data $\left\{ \left(
Q_{i},\delta _{i},T_{i},\mathbf{X}_{i}\right) \right\} _{i=1}^{n}$ into two
sub-samples given by different values of treatment status $T$; $\left\{
\left( Q_{i},\delta _{i},\mathbf{X}_{i}\right) \right\} _{i=1}^{n_{1}}$ are
those observations with $T_{i}=1\left( n_{1}=\sum_{i}T_{i}\right) $; and $%
\left\{ \left( Q_{i},\delta _{i},\mathbf{X}_{i}\right) \right\}
_{i=1}^{n_{0}}$ are those observations with $T_{i}=0\left(
n_{0}=\sum_{i}\left( 1-T_{i}\right) \right) $. Then, the task of estimating $%
F_{Q,\mathbf{X}|T}^{km}\left( y,\mathbf{x}|t\right) $ is reduced to
estimating (\ref{cumhazard}) and plugging it into (\ref{kmproduct}). We
estimate $\Lambda \left( y,\mathbf{x}|t\right) $ by replacing $H_{Q,\mathbf{X%
}|T}^{1}(y,\mathbf{x}|t)$ and $H_{Q,\mathbf{X}|T}(y-,\boldsymbol{\infty }|t)$
with their empirical analogues, leading to the estimator 
\begin{equation}
\hat{\Lambda}_{n}\left( y,\mathbf{x}|t\right) =\sum_{i=1}^{n_{t}}\frac{%
1\left\{ Q_{i:n_{t}}\leq y\right\} 1\left\{ \mathbf{X}_{\left[ i:n_{t}\right]
}\leq \mathbf{x}\right\} \delta _{\left[ i:n_{t}\right] }}{n_{t}-i+1},
\label{cumest}
\end{equation}%
where $Q_{1:n_{t}}\leq $ $\cdots \leq Q_{n_{t}:n_{t}}$ are the ordered $Q$%
-values in the sub-sample with $\left\{ T=t\right\} $, and $\mathbf{X}_{%
\left[ i:n_{t}\right] }$ and $\delta _{\left[ i:n_{t}\right] }$ are the
concomitants of the $i-th$ order statistic, that is, the $\mathbf{X}$ and $%
\delta $ paired with $Q_{i:n_{t}}$. Here, ties within outcomes of interest
or censoring random variables are ordered arbitrarily, and ties among $Y$
and $C$ are treated as if the former precedes the latter. By plugging $\hat{%
\Lambda}_{n}\left( y,\mathbf{x}|t\right) $ into (\ref{kmproduct}), and
noticing that $\hat{\Lambda}_{n}\left( y,\mathbf{x}|t\right) $ is a step
function, we have that a natural estimator for $F_{Q,\mathbf{X}%
|T}^{km}\left( y,\mathbf{x}|t\right) $ is 
\begin{equation}
\hat{F}_{Q,\mathbf{X}|T,n}^{km}\left( y,\mathbf{x}|t\right) =1-\dprod_{\bar{y%
}\leq y}\left[ 1-\hat{\Lambda}_{n}\left( \left\{ \bar{y}\right\} ,\mathbf{x}%
|t\right) \right] ,  \label{prodlim}
\end{equation}%
which is the multivariate extension of the time-honored \cite{Kaplan1958}
product limit estimator proposed by \cite{Stute1993}. As $\hat{F}_{Q,\mathbf{%
X}|T,n}^{km}\left( y,\mathbf{x}|t\right) $ is a step function, it can be
seen from (\ref{cumest}) and (\ref{prodlim}) that 
\begin{equation}
\hat{F}_{Q,\mathbf{X}|T,n}^{km}\left( y,\mathbf{x}|t\right)
=\sum_{i=1}^{n_{t}}W_{in_{t}}1\left\{ Q_{i:n_{t}}\leq y\right\} 1\left\{ 
\mathbf{X}_{\left[ i:n_{t}\right] }\leq \mathbf{x}\right\} ,  \label{kmsum}
\end{equation}%
where, for $1\leq i\leq n_{t},$ 
\begin{equation*}
W_{in_{t}}=\frac{\delta _{\lbrack i:n_{t}]}}{n_{t}-i+1}\prod_{j=1}^{i-1}%
\left[ \frac{n_{t}-j}{n_{t}-j+1}\right] ^{\delta _{\left[ j:n_{t}\right] }}
\end{equation*}%
is the Kaplan-Meier weight attached to $Q_{i:n_{t}},$ $t\in \left\{
0,1\right\} $.

Finally, given the discrete nature of $T$, we can nonparametrically estimate 
$\mathbb{P}\left( T=t\right) $ by its relative frequency $n_{t}/n.$ Putting
all these pieces together, we have that 
\begin{equation}
\hat{I}_{w,n}\left( y,\mathbf{x}\right) =\hat{I}_{w,n}^{1}\left( y,\mathbf{x}%
\right) -\hat{I}_{w,n}^{0}\left( y,\mathbf{x}\right) ,  \label{i0n}
\end{equation}%
where, for $t\in \left\{ 0,1\right\} $, 
\begin{equation}
\hat{I}_{w,n}^{t}\left( y,\mathbf{x}\right) =\frac{n_{t}}{n}%
\sum_{i=1}^{n_{t}}W_{in_{t}}\frac{1\left\{ Q_{i:n_{t}}\leq y\right\} w\left( 
\mathbf{X}_{\left[ i:n_{t}\right] },\mathbf{x}\right) }{\mathbb{\hat{P}}%
_{n}\left( T=t|\mathbf{X}_{\left[ i:n_{t}\right] }\right) }.  \label{id}
\end{equation}%
In the absence of censoring, $W_{in_{t}}=n_{t}^{-1}~a.s.$, and (\ref{id}) is
reduced to the empirical analogue of (\ref{Ieq}). Thus, it is evident that
our proposal is suitable for both censored and uncensored outcomes.

With $\hat{I}_{w,n}\left( y,\mathbf{x}\right) $ at hand, testing the null
hypothesis (\ref{cste}) is relatively straightforward: for a given weighting
function $w\left( \cdot ,\mathbf{x}\right) $, we just need to compare how
close $\sqrt{n}\hat{I}_{w,n}\left( y,\mathbf{x}\right) $ is to zero. We
consider the usual $sup$ and $L_{2}$ norms, with the indicator weighting
function $w\left( \mathbf{X},\mathbf{x}\right) =1\left\{ \mathbf{X}\leq 
\mathbf{x}\right\} $, leading to the Kolmogorov-Smirnov ($KS$) and Cram\'{e}%
r-von Mises ($CvM$) test statistics%
\begin{align}
KS_{n}& =\sqrt{n}\sup_{\left( y,\mathbf{x}\right) \in \mathcal{W}}\left\vert 
\hat{I}_{1,n}(y,\mathbf{x})\right\vert ,  \label{KS-0} \\
CvM_{n}& =n\int_{\mathcal{W}}\left\vert \hat{I}_{1,n}\left( y,\mathbf{x}%
\right) \right\vert ^{2}\hat{H}_{n}\left( dy,d\mathbf{x}\right) ,
\label{CvM-0}
\end{align}%
respectively, where $\hat{I}_{1,n}\left( y,\mathbf{x}\right) $ is defined as 
$\hat{I}_{w,n}\left( y,\mathbf{x}\right) $ with $w\left( \mathbf{X},\mathbf{x%
}\right) =1\left\{ \mathbf{X}\leq \mathbf{x}\right\} $, $\hat{H}_{n}\left( y,%
\mathbf{x}\right) $ denotes the sample analog of $H\left( y,\mathbf{x}%
\right) =\mathbb{P}\left( Q\leq y,\mathbf{X}\leq \mathbf{x}\right) $, and $%
\mathcal{W}=[-\infty ,\tau ]\times \mathcal{X}_{X}.$ Obviously, different
test statistics could be developed by applying other distances, or choosing
alternative weighting functions $w$, but for ease of exposition, we
concentrate of $KS_{n}$ and $CvM_{n}$. To avoid cumbersome notation, in the
rest of the article, we consider $\mathcal{W}=[-\infty ,\tau ]\times $ $%
\mathcal{X}_{X}$.

\begin{Rem}
\label{sub}In some circumstances, researchers may be interested in assessing
if the DTE is equal to zero for every subpopulation defined by a subset of $%
\mathbf{X}$, say $\mathbf{X}_{1}$. This is particularly attractive when the
dimension of $\mathbf{X}$ is moderate so one can plausibly rely on the
unconfoundedness assumption, but the covariates of interest for analyzing
heterogeneity are of much lower dimension; see e.g. \cite{Abrevaya2015} and 
\cite{Lee2017}. In this situation, instead of testing for (\ref{cste}), the
goal would be testing the null%
\begin{equation*}
H_{0}^{^{sub}}:F_{Y\left( 1\right) |\mathbf{X}_{1}}\left( y|\mathbf{\cdot }%
\right) -F_{Y\left( 0\right) |\mathbf{X}_{1}}\left( y|\mathbf{\cdot }\right)
=0~a.s.\text{ }\forall y\in \lbrack \mathcal{-\infty },\tau ].
\end{equation*}%
Note that by setting $w\left( \mathbf{X},\mathbf{x}\right) =1\left( \mathbf{X%
}_{1}\leq \mathbf{x}_{1}\right) $, or more generally, $w\left( \mathbf{X},%
\mathbf{x}\right) =w\left( \mathbf{X}_{1},\mathbf{x}_{1}\right) $, our tests
can cover this type of hypothesis in a rather straightforward manner.
\end{Rem}

\begin{Rem}
\label{one-sided}In some applications, researchers may be comfortable
assuming that the treatment effect cannot be \textquotedblleft
negative\textquotedblright\ for any subgroup. In such cases, one may want to
test%
\begin{equation*}
H_{0}:\Upsilon \left( y|\mathbf{X}\right) =0\text{ }a.s.~\forall y\in
\lbrack \mathcal{-\infty },\tau ],
\end{equation*}%
against the one-sided alternative hypothesis 
\begin{equation*}
H_{1}:\mathbb{P}\left( \Upsilon \left( y|\mathbf{X}\right) >0\right) >0\text{
for some }y\in \lbrack \mathcal{-\infty },\tau ],
\end{equation*}%
using the test statistic 
\begin{equation}
KS_{n}^{^{one}}=\sup_{\left( y,\mathbf{x}\right) \in \mathcal{W}}\sqrt{n}%
\hat{I}_{n}(t,\mathbf{x}).  \label{ks-one}
\end{equation}%
The statistical analysis of one-sided test based on (\ref{ks-one}) follows
similar arguments as the two-sided test based on (\ref{KS-0}).

We note that in settings without censoring, one can consider more general
tests for conditional stochastic dominance that do not rely on external
information about the direction of the (distributional) treatment effect,
see e.g. \cite{Andrews2013, Andrews2017} and \cite{Hsu2013}. Extending the
results in \cite{Andrews2013, Andrews2017} and \cite{Hsu2013} to allow for
random censoring, however, involves very technical challenges related to
establishing a uniform in DGP representation for Kaplan-Meier processes and
is, therefore, beyond the scope of this paper; see the Conclusion Section
for further details.
\end{Rem}

\section{Asymptotic Theory\label{Asymptotic-theory}}

\subsection{Asymptotic linear representation\label{km-int}}

We now discuss the asymptotic theory for our test statistics using the
following notation. For $t$ $\in \left\{ 0,1\right\} $, let $H_{t}\left(
y\right) =\mathbb{P}\left( Q\leq y,T=t\right) $, $H_{t,0}\left( y\right) =%
\mathbb{P}\left( Q\leq y,\delta =0,T=t\right) $, and $H_{t,11}\left(
y,x\right) =\mathbb{P}\left( Q\leq y,\mathbf{X}\leq \mathbf{x},\delta
=1,T=t\right) $. Define%
\begin{equation}
\gamma _{t,0}\left( \bar{y}\right) =\exp \left\{ \int_{0}^{\bar{y}-}\frac{%
H_{t,0}\left( d\bar{w}\right) }{1-H_{t}\left( \bar{w}\right) }\right\} .
\label{gam0}
\end{equation}%
Let%
\begin{equation}
\gamma _{t,1}\left( \bar{y};y,\mathbf{x}\right) =\frac{1}{1-H_{t}\left( \bar{%
y}\right) }\int 1\left\{ \bar{y}<\bar{w}\right\} \xi _{t}\left( \bar{w},%
\mathbf{\bar{x}},t;y,\mathbf{x}\right) \gamma _{t,0}\left( \bar{w}\right)
H_{t,11}\left( d\bar{w},d\mathbf{\bar{x}}\right)  \label{gam1}
\end{equation}%
and%
\begin{equation}
\gamma _{t,2}\left( \bar{y};y,\mathbf{x}\right) =\int \int \frac{1\left\{ 
\bar{v}<\bar{y},\bar{v}<\bar{w}\right\} \xi _{t}\left( \bar{w},\mathbf{\bar{x%
}},t;y,\mathbf{x}\right) }{\left[ 1-H_{t}\left( \bar{v}\right) \right] ^{2}}%
\gamma _{t,0}\left( \bar{w}\right) H_{t,0}\left( d\bar{v}\right)
H_{t,11}\left( d\bar{w},d\mathbf{\bar{x}}\right) ,  \label{gam2}
\end{equation}%
where%
\begin{eqnarray}
\xi _{1}\left( Q,\mathbf{X},T;y,\mathbf{x}\right) &=&\frac{T1\left\{ Q\leq
y\right\} 1\left\{ \mathbf{X}\leq \mathbf{x}\right\} }{p_{0}\left( \mathbf{X}%
\right) },  \label{xsi1} \\
\xi _{0}\left( Q,\mathbf{X},T;y,\mathbf{x}\right) &=&\frac{\left( 1-T\right)
1\left\{ Q\leq y\right\} 1\left\{ \mathbf{X}\leq \mathbf{x}\right\} }{%
1-p_{0}\left( \mathbf{X}\right) }.  \label{xsi0}
\end{eqnarray}%
Put 
\begin{equation}
\eta _{t,i}\left( y,\mathbf{x}\right) =\xi _{t}\left( Q_{i},\mathbf{X}%
_{i},T_{i};y,\mathbf{x}\right) \gamma _{t,0}\left( Q_{i}\right) \delta
_{i}+\gamma _{t,1}\left( Q_{i};y,\mathbf{x}\right) \left( 1-\delta
_{i}\right) -\gamma _{t,2}\left( Q_{i};y,\mathbf{x}\right) .  \label{eta}
\end{equation}

Some remarks are necessary. First, (\ref{eta}) relies only on the
\textquotedblleft known\textquotedblright\ functions $\xi _{t}$, $t\in
\left\{ 0,1\right\} $. Then, as discussed in \cite{Stute1995a,Stute1996a},
the first term of $\eta _{t,i}\left( y,\mathbf{x}\right) $ has expectation $%
\mathbb{E}\left[ \xi _{t}\left( Q,\mathbf{X},T;y,\mathbf{x}\right) \right] $%
. The second and third terms have identical expectations and appear because
of the censoring. As is expected and desired, in the absence of censoring, $%
\gamma _{t,0}\left( \cdot \right) =1$ $a.s.$, and $\gamma _{t,1}\left( \cdot
\right) =\gamma _{t,2}\left( \cdot \right) =0$ $a.s.$.

Given that $\hat{I}_{1,n}(\cdot ,\cdot )$ is the difference of two empirical
KM integrals, define 
\begin{equation}
\eta _{i}\left( y,\mathbf{x}\right) =\eta _{1,i}\left( y,\mathbf{x}\right)
-\eta _{0,i}\left( y,\mathbf{x}\right) ,  \label{iid_rep}
\end{equation}%
the difference of (\ref{eta}) between the treated and control group.

To discuss the estimation effect coming from not knowing $p_{0}\left( \cdot
\right) $ in the KM-integrals, let%
\begin{equation}
\alpha _{1}\left( \mathbf{X};y,\mathbf{x}\right) =-\frac{F_{Y\left( 1\right)
|\mathbf{X}}\left( y|\mathbf{X}\right) 1\left\{ \mathbf{X}\leq \mathbf{x}%
\right\} }{p_{0}\left( \mathbf{X}\right) },~\alpha _{0}\left( \mathbf{X};y,%
\mathbf{x}\right) =\frac{F_{Y\left( 0\right) |\mathbf{X}}\left( y|\mathbf{X}%
\right) 1\left\{ \mathbf{X}\leq \mathbf{x}\right\} }{1-p_{0}\left( \mathbf{X}%
\right) }  \label{alpha1}
\end{equation}%
Notice that $\alpha _{1}\left( \cdot ;y,\mathbf{x}\right) $ and $\alpha
_{0}\left( \cdot ;y,\mathbf{x}\right) $ are nothing more than the
conditional expectation of the (functional) derivative of (\ref{xsi1}) and (%
\ref{xsi0}) with respect to $p_{0}(\cdot )$, respectively. Similarly to (\ref%
{iid_rep}), define 
\begin{equation}
\alpha \left( \mathbf{X};y,\mathbf{x}\right) =\alpha _{1}\left( \mathbf{X}%
;y,x\right) -\alpha _{0}\left( \mathbf{X};y,\mathbf{x}\right) .
\label{est-effect-p}
\end{equation}

In order to present our asymptotic results, we need to assume some
additional regularity conditions related to the estimation of the propensity
score $p_{0}\left( \cdot \right) $, and some integrability conditions to
guarantee that the variance of our test statistics is finite and that the
censoring effects do not dominate in the right tail. These technical
assumptions are stated in Appendix \ref{App-Assumption}.

\begin{lemma}
\label{uni_rep}Under Assumptions \ref{unconfound}-\ref%
{censoring_identification} and Assumptions \ref{support_pscore_series}-\ref%
{moments_dte} stated in Appendix \ref{App-Assumption}, we have%
\begin{equation*}
\sqrt{n}\left( \hat{I}_{1,n}\left( y,\mathbf{x}\right) -I_{1}\left( y,%
\mathbf{x}\right) \right) =\frac{1}{\sqrt{n}}\sum_{i=1}^{n}\left\{ \left[
\eta _{i}\left( y,\mathbf{x}\right) -I_{1}\left( y,\mathbf{x}\right) \right]
+\alpha \left( \mathbf{X}_{i};y,\mathbf{x}\right) \left( T_{i}-p_{0}(\mathbf{%
X}_{i})\right) \right\} +o_{\mathbb{P}}\left( 1\right)
\end{equation*}%
uniformly in $\left( y,\mathbf{x}\right) \in \mathcal{W}.$
\end{lemma}

\subsection{Asymptotic null distribution}

Using the uniform representation from Lemma \ref{uni_rep}, we next establish
the weak convergence of the processes $\sqrt{n}\hat{I}_{1,n}\left( y,\mathbf{%
x}\right) $ under the null hypothesis 
\begin{equation}
H_{0}:\Upsilon \left( y|\mathbf{X}\right) =0~a.s.~\forall y\in \lbrack
-\infty ,\tau ].  \label{cstecens}
\end{equation}

\begin{theorem}
\label{null_0_DTE}Under the null hypothesis (\ref{cstecens}), Assumptions %
\ref{unconfound}-\ref{censoring_identification} and Assumptions \ref%
{support_pscore_series}-\ref{moments_dte} stated in Appendix \ref%
{App-Assumption}, we have%
\begin{equation*}
\sqrt{n}\hat{I}_{1,n}\left( y,\mathbf{x}\right) \Rightarrow C_{\infty
}\left( y,\mathbf{x}\right) \text{,}
\end{equation*}%
where $C_{\infty }$ is Gaussian process with zero mean and covariance
function%
\begin{equation}
V\left( \left( y_{1},\mathbf{x}_{1}\right) ,\left( y_{2},\mathbf{x}%
_{2}\right) \right) =\mathbb{E}\left[ \psi \left( y_{1},\mathbf{x}%
_{1}\right) \psi \left( y_{2},\mathbf{x}_{2}\right) \right] ,
\label{Variance}
\end{equation}%
with $\psi \left( y,\mathbf{x}\right) =\eta \left( y,\mathbf{x}\right)
+\alpha \left( \mathbf{X};y,\mathbf{x}\right) \left( T-p(\mathbf{X})\right)
. $
\end{theorem}

Now, we can apply the continuous mapping theorem to characterize the
limiting null distribution\ of our test statistics using the $\sup $ and $%
L_{2}$ distances.

\begin{corollary}
\label{test_DTE_0} Under the null hypothesis (\ref{cstecens}) and the
Assumptions of Theorem \ref{null_0_DTE},%
\begin{align*}
& KS_{n}\overset{d}{\rightarrow }\sup_{\left( y,\mathbf{x}\right) \in 
\mathcal{W}}\left\vert C_{\infty }\left( y,\mathbf{x}\right) \right\vert , \\
& CvM_{n}\overset{d}{\rightarrow }\int_{\mathcal{W}}\left\vert C_{\infty
}\left( y,\mathbf{x}\right) \right\vert ^{2}H\left( dy,d\mathbf{x}\right) .
\end{align*}
\end{corollary}

Let $T_{n}$ be a generic notation for $KS_{n}$ and $CvM_{n}$. From Corollary %
\ref{test_DTE_0}, it follows immediately that 
\begin{equation*}
\lim_{n\rightarrow \infty }\mathbb{P}\left\{ T_{n}>c_{\alpha
}^{^{T}}\right\} =\alpha
\end{equation*}%
where $c_{\alpha }^{^{T}}=\inf \left\{ c\in \lbrack 0,\infty
):\lim_{n\rightarrow \infty }\mathbb{P}\left\{ T_{n}>c\right\} =\alpha
\right\} .$

\subsection{Asymptotic power against fixed and local alternatives}

Now we analyze the asymptotic properties of our tests under the fixed
alternative $H_{1}$. Under $H_{1}$, $\mathbb{P}\left( \Upsilon \left( y|%
\mathbf{X}\right) =0\right) <1$ for some $y\in \lbrack -\infty ,\tau ]$,
implying that $I_{1}\left( y,\mathbf{x}\right) \not=0$ for some $\left( y,%
\mathbf{x}\right) \in \mathcal{W}$. Therefore,\ our test statistics $KS_{n}$
and $CvM_{n}$ diverge to infinity. Given that the critical values are
bounded, it follows that our tests are consistent. We formalize this result
in the next theorem.

\begin{theorem}
\label{H1_DTE}Under the fixed alternative hypothesis 
\begin{equation*}
H_{1}:\mathbb{P}\left( \Upsilon \left( y|\mathbf{X}\right) =0\right) <1\text{
for some }y\in \lbrack -\infty ,\tau ],
\end{equation*}%
and the Assumptions of Theorem \ref{null_0_DTE}, 
\begin{align*}
\lim_{n\rightarrow \infty }\mathbb{P}\left\{ KS_{n}>c_{\alpha
}^{^{KS}}\right\} & =1, \\
\lim_{n\rightarrow \infty }\mathbb{P}\left\{ CvM_{n}>c_{\alpha
}^{^{CvM}}\right\} & =1.
\end{align*}
\end{theorem}

Given that our test statistics diverge to infinity under fixed alternatives,
it is desirable studying the asymptotic power of these tests under local
alternatives. To this end, we study the asymptotic behavior of $\hat{I}%
_{1,n}\left( y,\mathbf{x}\right) $ under alternative hypotheses converging
to the null at the parametric rate $n^{-1/2}$.

Consider the following class of local alternatives:%
\begin{equation}
H_{1,n}:\Upsilon \left( y|\mathbf{X}\right) =\frac{h\left( y,\mathbf{X}%
\right) }{\sqrt{n}}\text{ }a.s.\text{ }\forall y\in \lbrack -\infty ,\tau ].
\label{cste_local}
\end{equation}

In the sequel, we need that (\ref{cste_local}) satisfies the following
regularity condition.

\begin{assumption}
\label{local alternative}

($a$) $h\left( \cdot ,\cdot \right) $ is an $F$-integrable function;

($b$) the set $h_{n}\equiv \left[ \left( y,\mathbf{x}\right) \in \mathcal{W}%
:n^{-1/2}h\left( y,\mathbf{x}\right) \not=0\right] $ has positive Lebesgue
measure.
\end{assumption}

\begin{theorem}
\label{local_cste}Under local alternatives (\ref{cste_local}), Assumptions %
\ref{unconfound}-\ref{censoring_identification}, \ref{local alternative} and
Assumptions \ref{support_pscore_series}-\ref{moments_dte} stated in Appendix %
\ref{App-Assumption},%
\begin{equation*}
\sqrt{n}\hat{I}_{1,n}\left( y,\mathbf{x}\right) \Rightarrow C_{\infty
}\left( y,\mathbf{x}\right) +R\left( y,\mathbf{x}\right)
\end{equation*}%
where $C_{\infty }$ is the process defined in Theorem \ref{null_0_DTE} and $%
R\left( y,\mathbf{x}\right) \equiv \mathbb{E}\left[ h\left( y,\mathbf{X}%
\right) 1\left\{ \mathbf{X}\leq \mathbf{x}\right\} \right] .$
\end{theorem}

From the above theorem and straightforward application of the continuous
mapping theorem, we see that our test statistics, under local alternatives
of the form of (\ref{cste_local}), converge to a different distribution
because of the presence of a deterministic shift function $R$. This
additional term guarantees the good local power property of our tests.

\subsection{Estimation of critical values\label{Bootstrap}}

From the above theorems, we see that the asymptotic distribution of $\sqrt{n}%
\hat{I}_{1,n}\left( \cdot ,\cdot \right) $ depends on the underlying data
generating process and standardization is complicated. To overcome this
problem, we propose computing critical values with the assistance of a
multiplier bootstrap. The proposed procedure has good theoretical and
empirical properties, is straightforward to verify its asymptotic validity,
is computationally easy to implement and does not require computing new
parameter estimates at each bootstrap replication.

To implement the bootstrap, we need nonparametric estimators for all the
terms in the asymptotic linear representation of Lemma \ref{uni_rep}, namely
the propensity score $p_{0}\left( \cdot \right) $, $\eta \left( y,\mathbf{x}%
\right) $ as in (\ref{iid_rep}), and $\alpha \left( \cdot ;y,\mathbf{x}%
\right) $ as in (\ref{est-effect-p}).

As already discussed, we estimate $p_{0}\left( \cdot \right) $ using the SLE
of \cite{Hirano2003}. In order to estimate $\eta \left( y,\mathbf{x}\right) $%
, we notice that after plugging in $\hat{p}_{n}\left( \cdot \right) $, each $%
\gamma $ only depends on $H$-functions and is, therefore, estimable by just
replacing the $H$-terms by their empirical counterparts. Then, we estimate $%
\eta \left( y,\mathbf{x}\right) $ by its empirical analogue%
\begin{equation*}
\hat{\eta}_{n}\left( y,\mathbf{x}\right) =\hat{\eta}_{1,n}\left( y,\mathbf{x}%
\right) -\hat{\eta}_{0,n}\left( y,\mathbf{x}\right) ,
\end{equation*}%
where, for $t\in \left\{ 0,1\right\} ,$%
\begin{align*}
\hat{\eta}_{t,n}\left( y,\mathbf{x}\right) & =\hat{\xi}_{t,n}\left( Q,%
\mathbf{X},T;y,\mathbf{x}\right) \hat{\gamma}_{t,0,n}\left( Q\right) \delta
_{t}+\hat{\gamma}_{t,1,n}\left( Q\right) \left( 1-\delta \right) -\hat{\gamma%
}_{t,2,n}\left( Q\right) ,\allowbreak \\
\hat{\gamma}_{t,0,n}\left( \bar{y}\right) & =\exp \left\{ \int_{0}^{\bar{y}-}%
\frac{\hat{H}_{t,0,n}\left( d\bar{w}\right) }{1-\hat{H}_{t,n}\left( \bar{w}%
\right) }\right\} ,\allowbreak \\
\hat{\gamma}_{t,1,n}\left( \bar{y}\right) & =\frac{1}{1-\hat{H}_{t,n}\left( 
\bar{y}\right) }\int 1\left\{ \bar{y}<\bar{w}\right\} \hat{\xi}_{t,n}\left( 
\bar{w},\mathbf{\bar{x}},t;y,\mathbf{x}\right) \hat{\gamma}_{t,0,n}\left( 
\bar{w}\right) \hat{H}_{t,11,n}\left( d\bar{w},d\mathbf{\bar{x}}\right)
,\allowbreak \\
\hat{\gamma}_{t,2,n}\left( \bar{y}\right) & =\int \int \frac{1\left\{ \bar{v}%
<\bar{y},\bar{v}<\bar{w}\right\} \hat{\xi}_{t,n}\left( \bar{w},\mathbf{\bar{x%
}},t;y,\mathbf{x}\right) }{\left[ 1-\hat{H}_{t,n}\left( \bar{v}\right) %
\right] ^{2}}\hat{\gamma}_{t,0,n}\left( \bar{w}\right) \hat{H}_{t,0,n}\left(
d\bar{v}\right) \hat{H}_{t,11,n}\left( d\bar{w},d\mathbf{\bar{x}}\right) ,
\end{align*}%
\linebreak where $\hat{\xi}_{1,n}\left( \cdot ,\cdot ,\cdot ;y,\mathbf{x}%
\right) $ and $\hat{\xi}_{0,n}\left( \cdot ,\cdot ,\cdot ;y,\mathbf{x}%
\right) $ are defined as in (\ref{xsi1}) and (\ref{xsi0}), respectively, but
with the true propensity score $p_{0}\left( \cdot \right) $ replaced by its
SLE $\hat{p}_{n}\left( \cdot \right) $, and 
\begin{align*}
\hat{H}_{t,n}\left( \bar{w}\right) & =\frac{1}{n}\sum_{i=1}^{n}1\left\{
Q_{i}\leq \bar{w}\right\} 1\left\{ T_{i}=t\right\} , \\
\hat{H}_{t,0,n}\left( \bar{w}\right) & =\frac{1}{n}\sum_{i=1}^{n}\left(
1-\delta _{i}\right) 1\left\{ Q_{i}\leq \bar{w}\right\} 1\left\{
T_{i}=t\right\} , \\
\hat{H}_{t,11,n}\left( \bar{w},\mathbf{\bar{x}}\right) & =\frac{1}{n}%
\sum_{i=1}^{n}\delta _{i}1\left\{ Q_{i}\leq \bar{w}\right\} 1\left\{ \mathbf{%
X}_{i}\leq \mathbf{\bar{x}}\right\} 1\left\{ T_{i}=t\right\} ,
\end{align*}%
are the sample counterparts of $H_{t}\left( \bar{w}\right) $, $H_{t,0}\left( 
\bar{w}\right) $ and $H_{t,11}\left( \bar{w},\mathbf{\bar{x}}\right) $,
respectively.

Finally, we must consider nonparametric estimators for $\alpha \left( \cdot
;y,\mathbf{x}\right) =$ $\alpha _{1}\left( \cdot ;y,\mathbf{x}\right)
-\alpha _{0}\left( \cdot ;y,\mathbf{x}\right) $, $\alpha _{1}\left( \cdot ;y,%
\mathbf{x}\right) $ and $\alpha _{1}\left( \cdot ;y,\mathbf{x}\right) $
being defined in (\ref{alpha1}). To this end, we must estimate $F_{Y\left(
0\right) |\mathbf{X}}\left( y|\mathbf{\cdot }\right) $ and $F_{Y\left(
1\right) |\mathbf{X}}\left( y|\mathbf{\cdot }\right) $. In the absence of
censored data, \cite{Donald2013} propose estimating these functions using
nonparametric series regression. Given that the outcome of interest $Y$ is
subjected to censoring, such a procedure is not at our disposal.
Notwithstanding, by using the Kaplan-Meier weights as discussed in Sections %
\ref{Baseline} and \ref{km-int}, we can overcome such a problem and estimate 
$F_{Y\left( 0\right) |\mathbf{X}}\left( y|\mathbf{x}\right) $ and $%
F_{Y\left( 1\right) |\mathbf{X}}\left( y|\mathbf{x}\right) $ by the
Kaplan-Meier series estimators:%
\begin{multline}
\hat{F}_{Y\left( 0\right) |\mathbf{X},n}^{km}\left( y|\mathbf{x}\right)
=\left( \frac{n_{0}}{n}\sum_{i=1}^{n_{0}}W_{in_{0}}\frac{1\left\{ Q_{\left[
i:n_{0}\right] }\leq y\right\} }{1-\hat{p}_{n}\left( \mathbf{X}_{\left[
i:n_{0}\right] }\right) }\mathbf{R}^{L}\left( \mathbf{X}_{\left[ i:n_{0}%
\right] }\right) \right) ^{\prime }  \label{kmseries0} \\
\left( \frac{1}{n}\sum_{i=1}^{n}\mathbf{R}^{L}\left( \mathbf{X}_{i}\right) 
\mathbf{R}^{L}\left( \mathbf{X}_{i}\right) ^{\prime }\right) ^{-1}\mathbf{R}%
^{L}\left( \mathbf{x}\right) ,
\end{multline}%
and%
\begin{multline}
\hat{F}_{Y\left( 1\right) |\mathbf{X},n}^{km}\left( y|\mathbf{x}\right)
=\left( \frac{n_{1}}{n}\sum_{i=1}^{n_{1}}W_{in_{1}}\frac{1\left\{ Q_{\left[
i:n_{1}\right] }\leq y\right\} }{\hat{p}_{n}\left( \mathbf{X}_{\left[ i:n_{1}%
\right] }\right) }\mathbf{R}^{L}\left( \mathbf{X}_{\left[ i:n_{1}\right]
}\right) \right) ^{\prime }  \label{kmseries1} \\
\left( \frac{1}{n}\sum_{i=1}^{n}\mathbf{R}^{L}\left( \mathbf{X}_{i}\right) 
\mathbf{R}^{L}\left( \mathbf{X}_{i}\right) ^{\prime }\right) ^{-1}\mathbf{R}%
^{L}\left( \mathbf{x}\right) ,
\end{multline}%
where $\mathbf{R}^{L}\left( \cdot \right) $ is the same power series used in
the SLE estimator, with a potentially different number of series. Armed with
(\ref{kmseries0}) and (\ref{kmseries1}), we can estimate $\alpha \left(
\cdot ;y,\mathbf{x}\right) $ by 
\begin{equation*}
\hat{\alpha}_{n}^{km}\left( \mathbf{X};y,\mathbf{x}\right) =-\left( \frac{%
\hat{F}_{Y\left( 1\right) |\mathbf{X},n}^{km}\left( y|\mathbf{X}\right) }{%
\hat{p}_{n}\left( \mathbf{X}\right) }+\frac{\hat{F}_{Y\left( 0\right)
|X,n}^{km}\left( y|\mathbf{X}\right) }{1-\hat{p}_{n}\left( \mathbf{X}\right) 
}\right) 1\left\{ \mathbf{X}\leq \mathbf{x}\right\} .
\end{equation*}

Once we have nonparametric estimators for $p_{0}\left( \cdot \right) $, $%
\eta \left( y,\mathbf{x}\right) $, and $\alpha \left( \cdot ;y,\mathbf{x}%
\right) $, the bootstrapped version of $\hat{I}_{1,n}\left( y,\mathbf{x}%
\right) $ is given by 
\begin{equation}
\hat{I}_{1,n}^{\ast }\left( y,\mathbf{x}\right) {\small =}\frac{1}{n}%
\sum_{i=1}^{n}\left[ \hat{\eta}_{i,n}\left( y,\mathbf{x}\right) +\hat{\alpha}%
_{n}^{KM}\left( \mathbf{X}_{i};y,\mathbf{x}\right) \left( T_{i}-\hat{p}%
_{n}\left( \mathbf{X}_{i}\right) \right) \right] V_{i}  \label{boot-in}
\end{equation}%
where $\hat{\eta}_{n}\left( y,\mathbf{x}\right) =\hat{\eta}_{1,n}\left( y,%
\mathbf{x}\right) -\hat{\eta}_{0,n}\left( y,\mathbf{x}\right) $, and the
random variables $\left\{ V_{i}\right\} _{i=1}^{n}$ are $iid$ with bounded
support, zero mean and variance one, being independently generated from the
sample \linebreak $\left\{ \left( Q_{i},\delta _{i},T_{i},\mathbf{X}%
_{i}\right) \right\} _{i=1}^{n}$. A popular example involves $iid$ Bernoulli
variables $\left\{ V_{i}\right\} $ with \linebreak $\mathbb{P}\left(
V=1-\kappa \right) =\kappa /\sqrt{5}$ and $\mathbb{P}\left( V=\kappa \right)
=1-\kappa /\sqrt{5}$, where $\kappa =\left( \sqrt{5}+1\right) /2,$ as
suggested by \cite{Mammen1993}.

Replacing $\hat{I}_{1,n}\left( y,\mathbf{x}\right) $ with $\hat{I}%
_{1,n}^{^{\ast }}\left( y,\mathbf{x}\right) ,$ we get the bootstrap versions
of $KS_{n}$ and $CvM_{n}$, $KS_{n}^{^{\ast }}$ and $CvM_{n}^{^{\ast }}$,
respectively. The asymptotic critical values are estimated by%
\begin{align*}
c_{n,\alpha }^{^{KS,~\ast }}& \equiv \inf \left\{ c_{\alpha }\in \lbrack
0,\infty ):\lim_{n\rightarrow \infty }\mathbb{P}_{n}^{^{\ast }}\left\{
KS_{n}^{\ast }>c_{\alpha }\right\} =\alpha \text{ }\right\} , \\
c_{n,\alpha }^{^{CvM,~\ast }}& \equiv \inf \left\{ c_{\alpha }\in \lbrack
0,\infty ):\lim_{n\rightarrow \infty }\mathbb{P}_{n}^{^{\ast }}\left\{
CvM_{n}^{\ast }>c_{\alpha }\right\} =\alpha \text{ }\right\}
\end{align*}%
where $\mathbb{P}_{n}^{^{\ast }}$ means bootstrap probability, i.e.
conditional on the sample $\left\{ \left( Q_{i},\delta _{i},T_{i},\mathbf{X}%
_{i}\right) \right\} _{i=1}^{n}.$ In practice, $c_{n,\alpha }^{^{KS,~\ast }}$
and $c_{n,\alpha }^{^{CvM,~\ast }}$ are approximated as accurately as
desired by $\left( KS_{n}^{^{\ast }}\right) _{B(1-\alpha )}$ and $\left(
CvM_{n}^{^{\ast }}\right) _{B(1-\alpha )}$, the $B\left( 1-\alpha \right)
-th $ order statistic from $B$ replicates $\left\{ KS_{n}^{^{b,\ast
}}\right\} _{l=1}^{B}$ of $KS_{n}^{^{\ast }}$ or $\left\{ CvM_{n}^{^{b,\ast
}}\right\} _{l=1}^{B}$ of $CvM_{n}^{^{\ast }}$, respectively.

\begin{theorem}
\label{null_DTE}Let Assumptions \ref{unconfound}, \ref%
{censoring_identification}, \ref{local alternative} and Assumptions \ref%
{support_pscore_series}-\ref{moments_dte} stated in Appendix \ref%
{App-Assumption} hold. Assume $\left\{ V_{i}\right\} _{i=1}^{n}$ are $iid$,
independent of the sample $\left\{ \left( Q_{i},\delta _{i},T_{i},\mathbf{X}%
_{i}\right) \right\} _{i=1}^{n}$, bounded random variables with zero mean
and variance one. Then, under the null hypothesis (\ref{cste}), any fixed
alternative hypothesis, or under the local alternatives (\ref{cste_local}) 
\begin{equation*}
\sqrt{n}\hat{I}_{1,n}^{\ast }\left( y,\mathbf{x}\right) \underset{\ast }{%
\Rightarrow }C_{\infty }\left( y,\mathbf{x}\right) \text{\ }
\end{equation*}%
where $C_{\infty }$ is the same Gaussian process of Theorem \ref{null_0_DTE}
and $\underset{\ast }{\Rightarrow }$ denoting weak convergence in
probability under the the bootstrap law (see \cite{Gine1990}).
\end{theorem}

The following algorithm presents a complete procedure to implement our
bootstrap-based tests for the null hypothesis of zero conditional
distributional treatment effect as in (\ref{cstecens}). We focus on the Cram%
\'{e}r-von Mises test but the Kolmogorov-Smirnov test is formed analogously.

\begin{algorithm}
\begin{enumerate}
\item \label{alg:boot1} Construct an estimate for the propensity score using 
$\hat{p}_{n}\left( \mathbf{x}\right) =\mathcal{L}\left( \mathbf{R}^{L}\left(
x\right) ^{\prime }\boldsymbol{\hat{\pi}}_{L}\right) $, with $\boldsymbol{%
\hat{\pi}}_{L}$ as defined in (\ref{pscore-series}).

\item For all data points $\left( y,\mathbf{x}\right) =\left( Y_{i},\mathbf{X%
}_{i}\right) $, $i=1,\dots ,n$, compute $\hat{I}_{1,n}\left( y,\mathbf{x}%
\right) $ using (\ref{i0n}), and construct the test statistic $CvM_{n}$
statistics as in (\ref{CvM-0}).

\item Draw a realization of $\{V_{i}\}_{i=1}^{n}.$

\item Compute $\hat{I}_{1,n}^{\ast }\left( y,\mathbf{x}\right) $ as in (\ref%
{boot-in}) and form $CvM_{n}^{^{b,\ast }}$ statistic as (\ref{CvM-0}), but
with $\hat{I}_{1,n}^{\ast }\left( y,\mathbf{x}\right) $ playing the role of $%
\hat{I}_{1,n}\left( y,\mathbf{x}\right) $.

\item Repeat steps 3-4 $B$ times to obtain $\left\{ CvM_{n}^{^{b,\ast
}}\right\} _{b=1}^{B}$.

\item Construct $c_{n,\alpha }^{^{CvM,~\ast }}$ as the empirical $\left(
1-a\right) $-quantile of the $B$ bootstrap draws of $CvM_{n}^{^{b,\ast }}$.

\item If $CvM_{n}>c_{n,\alpha }^{^{CvM,~\ast }}$, reject the null hypothesis
(\ref{cstecens}). Otherwise, do not reject (\ref{cstecens}).
\end{enumerate}
\end{algorithm}

\section{Some extensions of the basic setup\label{Extensions}}

\subsection{Testing for the Zero Conditional Average Treatment Effect\label%
{ate-sec}}

So far, we have only discussed tests for the existence of distribution
treatment effects. Although the proposed tests for zero conditional
distribution treatment effect are able to detect a very broad set of
alternative hypotheses, they are still not able to pin down the direction of
the departure from the null. For instance, if we reject the null (\ref{cste}%
), we unfortunately do not know if the policy affects the conditional mean
or, instead, any other particular feature of the outcome distribution (e.g.,
its 5th moment). Being able to differentiate such cases is important: policy
makers may be in favor of implementing a job training that reduces the
average unemployment durations, but may be more reluctant to implement such
policy if there is evidence that it affects only the other higher order
moments. Given the major role played by the average treatment effect, in
this section we show how to adapt our DTE tests to focus on this particular
TE measure.

Let $\Upsilon ^{^{cate}}\left( \mathbf{X}\right) \equiv $ $\mathbb{E}\left[
Y\left( 1\right) |\mathbf{X}\right] -$ $\mathbb{E}\left[ Y\left( 0\right) |%
\mathbf{X}\right] $ be the conditional average treatment effect. From Lemma %
\ref{identification}, we have that identification of $\Upsilon
^{^{cate}}\left( \mathbf{\cdot }\right) $ is not guaranteed unless the
support of the censoring variable is larger than or equal to the support the
potential outcomes of interest. Given that in follow-up studies such
condition is usually violated, it may be more appropriate to focus on the
restricted conditional average treatment effect (CATE), 
\begin{equation*}
\Upsilon _{\bar{\tau}}^{^{cate}}\left( \mathbf{X}\right) \equiv \mathbb{E}%
\left[ Y\left( 1\right) 1\left\{ Y\left( 1\right) \leq \bar{\tau}\right\} |%
\mathbf{X}\right] -\mathbb{E}\left[ Y\left( 0\right) 1\left\{ Y\left(
0\right) \leq \bar{\tau}\right\} |\mathbf{X}\right] ,
\end{equation*}%
see e.g. \cite{Zucker1998}. From Lemma \ref{identification} we know that $%
\Upsilon _{\bar{\tau}}^{^{cate}}\left( \mathbf{\cdot }\right) $ is
nonparametrically identified for all $\bar{\tau}\leq \tau $.

We are concerned with the following hypothesis:%
\begin{equation}
H_{0}^{^{cate}}:\Upsilon _{\bar{\tau}}^{^{cate}}\left( \mathbf{X}\right) =0%
\text{ }a.s.  \label{h0-cate}
\end{equation}
Under $H_{0}^{^{cate}}$, the restricted average treatment effect (ATE) is
equal to zero for all subpopulations defined by covariates. The alternative
hypothesis $H_{1}^{^{cate}}$ is the negation of the null $H_{0}^{^{cate}}$.

From the same reasoning of Lemma \ref{integrated}, we can rewrite (\ref%
{h0-cate}) as 
\begin{equation*}
H_{0}^{^{cate}}:I_{\bar{\tau}}^{^{cate}}(\mathbf{x})=0~a.e.\text{ }in\text{ }%
\mathcal{X}_{X},
\end{equation*}%
where $I_{\bar{\tau}}^{^{cate}}(\mathbf{x})=I_{\bar{\tau}}^{^{1,cate}}\left( 
\mathbf{x}\right) -I_{\bar{\tau}}^{^{0}}\left( \mathbf{x}\right) $, with

\begin{equation*}
I_{\bar{\tau}}^{^{t,cate}}\left( \mathbf{x}\right) \equiv \mathbb{E}^{km}%
\left[ \dfrac{1\left\{ T=t\right\} Q1\left\{ Q\leq \bar{\tau}\right\} }{%
\mathbb{P}\left( T=t|\mathbf{X}\right) }1\left\{ \mathbf{X}\leq \mathbf{x}%
\right\} \right] ,~t\in \left\{ 0,1\right\} .
\end{equation*}%
Then, following the same steps as in Section \ref{teststat}, our $KS$ type
test statistic for hypothesis (\ref{h0-cate}) is%
\begin{equation*}
KS_{\bar{\tau},n}^{^{cate}}=\sup_{\mathbf{x}\in \mathcal{X}_{X}}\left\vert 
\sqrt{n}\hat{I}_{\bar{\tau},n}^{^{cate}}(\mathbf{x})\right\vert ,
\end{equation*}%
where $\hat{I}_{\bar{\tau},n}^{^{cate}}\left( y,\mathbf{x}\right) $ is
defined in (\ref{Icate}) at the Appendix \ref{App-B}. The discussion for the 
$CvM$ test is the same and is, therefore, omitted. Notice that when $\bar{%
\tau}=\tau $, $1\left\{ Q\leq \tau \right\} =1$ $a.s.$ and, therefore, no
user-chosen trimming is necessary. This is of particular importance because,
in this case, we are using all the information about the average treatment
effect available in the data. In the simulation and application in Sections %
\ref{MCsimul} and \ref{Illinois}, we follow exactly this convention.

Under similar conditions to those in Section \ref{Asymptotic-theory}, we can
derive the asymptotic linear representation of $\sqrt{n}\hat{I}_{\bar{\tau}%
,n}^{^{cate}}(\mathbf{x}).$ Using an analogous procedure to the one
described in Section \ref{Bootstrap}, let $c_{\bar{\tau},\alpha
,n}^{^{cate,\ast }}$ denote the bootstrap critical value of the $KS_{\bar{%
\tau},n}^{^{cate}}$.

\begin{theorem}
\label{null_tests}Suppose Assumptions \ref{unconfound}-\ref%
{censoring_identification} and Assumptions \ref{support_pscore_series}, \ref%
{derivatives_pscore}, \ref{series estimator}, \ref{diff-ate} and \ref%
{moments_ate} stated in Appendix \ref{App-Assumption} are satisfied. Then,
for a fixed $\bar{\tau}\leq \tau $,

\begin{enumerate}
\item Under $H_{0}^{^{cate}}$, $\lim_{n\rightarrow \infty }\mathbb{P}%
_{n}\left\{ KS_{\bar{\tau},n}^{^{cate}}>c_{\bar{\tau},\alpha
,n}^{^{cate,\ast }}\right\} =\alpha .$

\item Under $H_{1}^{^{cate}}$, $\lim_{n\rightarrow \infty }\mathbb{P}%
_{n}\left\{ KS_{\bar{\tau},n}^{^{cate}}>c_{\bar{\tau},\alpha
,n}^{^{cate,\ast }}\right\} =1.$

\item Under $H_{1,n}^{^{cate}}:\Upsilon _{\bar{\tau}}^{^{cate}}\left( 
\mathbf{X}\right) =n^{-1/2}h_{\bar{\tau}}^{^{cate}}\left( \mathbf{X}\right)
~a.s.$, if $h_{\bar{\tau}}^{^{cate}}\left( \cdot \right) $ is an integrable
function, and the set $h_{\bar{\tau},n}^{^{cate}}\equiv \left\{ \mathbf{x}%
\in \mathcal{X}_{X}:n^{-1/2}h_{\bar{\tau}}^{^{cate}}\left( \mathbf{x}\right)
\not=0\right\} $ has positive Lebesgue measure, then \linebreak $%
\lim_{n\rightarrow \infty }\mathbb{P}_{n}\left\{ KS_{\bar{\tau}%
,n}^{^{cate}}>c_{\bar{\tau},\alpha ,n}^{^{cate,\ast }}\right\} >\alpha .$
\end{enumerate}
\end{theorem}

The results in Theorem \ref{null_tests} are related to \cite{Crump2008}. In
the absence of censoring, \cite{Crump2008} propose a test for $%
H_{0}^{^{cate}}$ based on smooth estimates of the conditional average
treatment effect. In particular, they use a series approach to estimate $%
\mathbb{E}\left[ Y\left( 1\right) |\mathbf{X}\right] $ and $\mathbb{E}\left[
Y\left( 0\right) |\mathbf{X}\right] $ and then compare how close the smooth
estimate of $\Upsilon ^{^{cate}}\left( \cdot \right) $ is to zero. Given
that \cite{Crump2008} test is based on the \textquotedblleft local
approach\textquotedblright , their test for $H_{0}^{^{cate}}$ is not able to
detect local alternatives of the type of $H_{1,n}^{^{cate}}$ and may suffer
from the \textquotedblleft curse of dimensionality.\textquotedblright\ This
is in contrast with the results in Theorem \ref{null_tests}. Finally, note
that as highlighted in Remark \ref{sub}, our tests naturally cover the case
where researchers are interested in testing for zero CATE only for a subset
of covariates $\mathbf{X}_{1}$ of all available characteristics $\mathbf{X}$%
, whereas \cite{Crump2008} test does not. Thus, one can see that even when
censoring is not an issue, our proposal is attractive when compared to \cite%
{Crump2008}.

\subsection{Testing for Homogeneous Conditional Average Treatment Effect 
\label{hom-cate}}

In this section, we show how one can adapt our baseline framework to test
whether there is heterogeneity in the (restricted) ATE with respect to
observed characteristics. In simple terms, we want to assess whether
individuals with different background characteristics have different ATE.
Such hypothesis is particularly relevant for policy makers interested in
extending a pilot program to a larger population; if there is strong
evidence against the hypothesis of homogeneous effect, one may be more
concerned with targeting the appropriate population who should receive the
treatment, see e.g. \cite{Manski2004} and \cite{Crump2008}.

As in Section \ref{ate-sec}, we focus on the restricted CATE. We seek to
test 
\begin{equation}
H_{0}^{^{\hom }}:\exists ~\Upsilon _{\bar{\tau}}\in \mathbb{R}:\Upsilon _{%
\bar{\tau}}^{^{cate}}\left( \mathbf{X}\right) =\Upsilon _{\bar{\tau}}\text{ }%
a.s..  \label{cctae}
\end{equation}%
The alternative hypothesis $H_{1}^{^{\hom }}$ is the negation of $%
H_{0}^{^{\hom }}.$

Note that we can rewrite (\ref{cctae}) as%
\begin{equation*}
H_{0}^{^{\hom }}:I_{\bar{\tau}}^{^{\hom }}\left( \mathbf{x}\right) =0\text{ }%
a.e.\text{ in }\mathcal{X}_{X}
\end{equation*}%
where $I_{\bar{\tau}}^{^{\hom }}\left( \mathbf{x}\right) =I_{\bar{\tau}%
}^{^{1,\hom }}\left( \mathbf{x}\right) -I_{\bar{\tau}}^{^{0,\hom }}\left( 
\mathbf{x}\right) $,%
\begin{equation*}
I_{\bar{\tau}}^{^{t,\hom }}\left( \mathbf{x}\right) \equiv \mathbb{E}^{km}%
\left[ 1\left\{ T=t\right\} \left( \dfrac{Q1\left\{ Q\leq \bar{\tau}\right\} 
}{\mathbb{P}\left( T=t|\mathbf{X}\right) }-\left( 2T-1\right) I_{\bar{\tau}%
}^{^{ate}}\right) 1\left\{ \mathbf{X}\leq \mathbf{x}\right\} \right] ,
\end{equation*}%
$t\in \left\{ 0,1\right\} $, and $I_{\bar{\tau}}^{^{ate}}$ is the restricted
average treatment effect, 
\begin{equation*}
I_{\bar{\tau}}^{^{ate}}\equiv \mathbb{E}^{km}\left[ \dfrac{TQ1\left\{ Q\leq 
\bar{\tau}\right\} }{\mathbb{P}\left( T=1|\mathbf{X}\right) }\right] -%
\mathbb{E}^{km}\left[ \dfrac{\left( 1-T\right) Q1\left\{ Q\leq \bar{\tau}%
\right\} }{\mathbb{P}\left( T=0|\mathbf{X}\right) }\right] .
\end{equation*}%
Based on this characterization of $H_{0}^{^{\hom }}$, our proposed test
statistic for (\ref{cctae}) is%
\begin{equation*}
KS_{\bar{\tau},n}^{^{\hom }}=\sup_{\mathbf{x}\in \mathcal{X}_{X}}\left\vert 
\sqrt{n}\hat{I}_{\bar{\tau},n}^{^{\hom }}(\mathbf{x})\right\vert ,
\end{equation*}%
where $\hat{I}_{\bar{\tau},n}^{^{\hom }}\left( \mathbf{x}\right) $ is
defined in (\ref{Ihom}) in Appendix \ref{App-B}. Let $c_{\bar{\tau},\alpha
,n}^{^{\hom ,\ast }}$ denote the bootstrap critical value of the $KS_{\bar{%
\tau},n}^{^{\hom }}$.

\begin{theorem}
\label{tests_ccate}Suppose Assumptions \ref{unconfound}-\ref%
{censoring_identification} and Assumptions \ref{support_pscore_series}, \ref%
{derivatives_pscore}, \ref{series estimator}, \ref{diff-ate} and \ref%
{moments_ate} stated in Appendix \ref{App-Assumption} are satisfied. Then,
for a fixed $\bar{\tau}\leq \tau $,

\begin{enumerate}
\item Under $H_{0}^{^{\hom }}$, $\lim_{n\rightarrow \infty }\mathbb{P}%
_{n}\left\{ KS_{\bar{\tau},n}^{^{\hom }}>c_{\bar{\tau},\alpha ,n}^{^{\hom
,\ast }}\right\} =\alpha .$

\item Under $H_{1}^{^{\hom }}$, $\lim_{n\rightarrow \infty }\mathbb{P}%
_{n}\left\{ KS_{\bar{\tau},n}^{^{\hom }}>c_{\bar{\tau},\alpha ,n}^{^{\hom
,\ast }}\right\} =1.$

\item Under $H_{1,n}^{^{\hom }}:\Upsilon _{\bar{\tau}}\left( \mathbf{x}%
\right) -\Upsilon _{\bar{\tau}}=n^{-1/2}h_{\bar{\tau}}^{^{\hom }}\left( 
\mathbf{X}\right) ~a.s.$, if $h_{\bar{\tau}}^{^{\hom }}\left( \cdot \right) $
is an integrable function, and the set $h_{\bar{\tau},n}^{^{\hom }}\equiv
\left\{ \mathbf{x}\in \mathcal{X}_{X}:n^{-1/2}h_{\bar{\tau}}^{^{\hom
}}\left( \mathbf{x}\right) \not=0\right\} $ has positive Lebesgue measure,
then $\lim_{n\rightarrow \infty }\mathbb{P}_{n}\left\{ KS_{\bar{\tau}%
,n}^{^{\hom }}>c_{\bar{\tau},\alpha ,n}^{^{\hom ,\ast }}\right\} >\alpha $.
\end{enumerate}
\end{theorem}

The results in Theorem \ref{tests_ccate} are related to \cite{Crump2008},
who also proposed a test for $H_{0}^{^{\hom }}$ in a context in which
censoring is not present. The test in \cite{Crump2008} is not able to detect
local alternatives of the type of $H_{1,n}^{^{\hom }}$, and is not suitable
to assess the presence of ATE heterogeneity when the conditioning vector in (%
\ref{cctae}) is $\mathbf{X}_{1}$, a strict subset of $\mathbf{X}.$ As
discussed in Remark \ref{sub}, our test easily accommodates this situation.
Given these attractive features, we argue that, even when censoring is not
an issue, the results in Theorem \ref{tests_ccate} are of interest to
applied researchers and policy makers.

\subsection{Testing within the Local Treatment Effect setup\label{LATE}}

In many important applications, the assumption that treatment allocation is
exogenous may be too restrictive. For instance, when individuals do not
comply with their treatment assignment, or more generally when they sort
into treatment based on expected gains, Assumption \ref{unconfound} is
likely to be violated. The goal of this section is to show that, if the
unconfoundedness assumption does not hold, our tests are still applicable to
the local treatment effect (LTE) setup introduced by \cite{Imbens1994} and 
\cite{Angrist1996}.

The LTE setup presumes the availability of a binary instrumental variable $Z$
for the treatment assignment. Denote $T\left( 0\right) $ and $T\left(
1\right) $ the value that $T$ would have taken if $Z$ is equal to zero or
one, respectively. The realized treatment is $T=ZT\left( 1\right) +\left(
1-Z\right) T\left( 0\right) .$ Thus, the observed sample consists of $iid$
copies $\left\{ \left( Q_{i},\delta _{i},T_{i},Z_{i},\mathbf{X}_{i}\right)
\right\} _{i=1}^{n}$. Denote $q_{0}\left( \mathbf{X}\right) \equiv $ $%
\mathbb{P}(Z=1|\mathbf{X})$.

To identify the LTE for the subpopulation of compliers, that is, individuals
who comply with their actual assignment of treatment and would have complied
with the alternative assignment, we need the following assumptions.

\begin{assumption}
\label{late_a}($i$) $\left( Y\left( 0\right) ,Y\left( 1\right) ,T\left(
0\right) ,T\left( 1\right) ,C\left( 0\right) ,C\left( 1\right) \right) 
%TCIMACRO{\TeXButton{indep}{\independent}}%
%BeginExpansion
\independent%
%EndExpansion
Z|\mathbf{X}$; ($ii$) for some $\varepsilon >0$, $\varepsilon \leq
q_{0}\left( \mathbf{X}\right) \leq 1-\varepsilon $ $a.s.$ and $\mathbb{P}%
\left( T\left( 1\right) =1|\mathbf{X}\right) >\mathbb{P}\left( T\left(
0\right) =1|\mathbf{X}\right) $ $a.s.$; and ($iii$) $\mathbb{P}\left(
T\left( 1\right) \geq T\left( 0\right) |\mathbf{X}\right) \allowbreak =1$ $%
a.s.$.
\end{assumption}

\begin{assumption}
\label{censoring_identification_late}Assume that $(i)$ $\left( Y\left(
0\right) ,Y\left( 1\right) \right) $ $%
%TCIMACRO{\TeXButton{indep}{\independent}}%
%BeginExpansion
\independent%
%EndExpansion
\left( C\left( 0\right) ,C\left( 1\right) \right) |T,Z;$ and $(ii)$ for $%
t\in \left\{ 0,1\right\} $, $\mathbb{P}\left( Y\left( t\right) \leq C\left(
t\right) |\mathbf{X},T,Z,Y\left( t\right) \right) =\mathbb{P}\left( Y\left(
t\right) \leq C\left( t\right) |T,Z,Y\left( t\right) \right) $ $a.s.$.
\end{assumption}

Assumption \ref{late_a} is standard in the literature, see e.g. \cite%
{Abadie2002}, \cite{Abadie2003}, \cite{Frolich2007}. Assumption \ref%
{censoring_identification_late} is analogous to Assumption \ref%
{censoring_identification} and is necessary owing to the censoring. It is
important to notice that Assumption \ref{censoring_identification_late} does
not restrict how treatment status and instruments affect the censoring
variable, which is weaker than typical assumptions used in the literature,
see e.g. \cite{Frandsen2014}.

Because treatment effects are allowed to be arbitrarily heterogeneous, one
is only able to identify effects for the complier subpopulation, see e.g. 
\cite{Abadie2003}, \cite{Frandsen2014} and \cite{SantAnna2016}. Let $%
\Upsilon ^{^{ldte}}\left( y|\mathbf{X}\right) \equiv F_{Y\left( 1\right) |%
\mathbf{X}}\left( y|\mathbf{X},pop=comp\right) -F_{Y\left( 0\right) |\mathbf{%
X}}\left( y|\mathbf{X},pop=comp\right) $. Thus, our goal is to test the null
hypothesis%
\begin{equation}
H_{0}^{^{ldte}}:\Upsilon ^{^{ldte}}\left( y|\mathbf{X}\right) =0\text{ }a.s.%
\text{ }\forall y\mathbf{\in }[\mathcal{-\infty ,}\tau ],  \label{h0-ldte}
\end{equation}%
against $H_{1}^{^{ldte}}$, which is simply the negation of (\ref{h0-ldte}).
The null (\ref{h0-ldte}) is analogous to (\ref{cste}) within the LTE setup.
For conciseness, we concentrate our attention on $H_{0}^{^{ldte}}$, but of
course, we can also adapt the hypotheses discussed in\ Sections \ref{ate-sec}
and \ref{hom-cate} to the LTE setup in a routine fashion. Such extensions
are presented in the Supplementary Appendix.

In order to proceed, we must show that $\Upsilon ^{^{ldte}}\left( y|\mathbf{%
\cdot }\right) $ can be written in terms of observables $\left( Q,\delta
,T,Z,\mathbf{X}\right) $. In the Supplementary Appendix, we show that this
is the case by extending Lemma \ref{identification} to the LTE setup. Then,
using the integrated moment approach analogous to Lemma \ref{integrated}, we
can show that $H_{0}^{^{ldte}}$ is true if and only if%
\begin{equation*}
I^{^{ldte}}\left( y,\mathbf{x}\right) =0~a.e.\text{ in }[-\infty ,\tau
]\times \mathcal{X}_{X}\mathcal{,}
\end{equation*}%
where $I^{^{ldte}}\left( y,\mathbf{x}\right) =I^{^{1,ldte}}\left( y,\mathbf{x%
}\right) -I^{^{0,ldte}}\left( y,\mathbf{x}\right) $, and for $t\in \left\{
0,1\right\} $,%
\begin{multline*}
I^{^{t,ldte}}\left( y,\mathbf{x}\right) \equiv \left( 2t-1\right) \left\{ 
\mathbb{E}^{km}\left[ \frac{1\left\{ Q\leq y\right\} }{q_{0}\left( \mathbf{X}%
\right) }1\left\{ \mathbf{X}\leq \mathbf{x}\right\} |T=t,Z=1\right] \mathbb{P%
}\left( T=t,Z=1\right) \right. \\
\left. -\mathbb{E}^{km}\left[ \frac{1\left\{ Q\leq y\right\} }{1-q_{0}\left( 
\mathbf{X}\right) }1\left\{ \mathbf{X}\leq \mathbf{x}\right\} |T=t,Z=0\right]
\mathbb{P}\left( T=t,Z=0\right) \right\}
\end{multline*}%
Then, following the discussion in Section \ref{teststat}, our $KS$ type test
statistic for hypothesis (\ref{h0-ldte}) is%
\begin{equation*}
KS_{n}^{^{ldte}}=\sqrt{n}\sup_{\left( y,\mathbf{x}\right) \in \mathcal{W}%
}\left\vert \hat{I}_{n}^{^{ldte}}(y,\mathbf{x})\right\vert ,
\end{equation*}%
where $\hat{I}_{n}^{^{ldte}}(y,\mathbf{x})$ is defined in (\ref{Ildte}) in
Appendix \ref{App-B}. Let $c_{\alpha ,n}^{^{ldte,\ast }}$ denote the
bootstrap critical value of the $KS_{n}^{^{ldte}}$.

\begin{theorem}
\label{tests_ldte}Suppose Assumptions \ref{late_a}-\ref%
{censoring_identification_late} are satisfied. Further, suppose that for the
subpopulation of compliers, Assumption \ref{support_pscore_series}, \ref%
{diff-dte} and \ref{moments_dte} stated in the Appendix are satisfied and
that $q_{0}$ and its SLE $\hat{q}_{n}$ satisfy the analogous of Assumptions %
\ref{derivatives_pscore} and \ref{series estimator}. Then,

\begin{enumerate}
\item Under $H_{0}^{^{ldte}}$, $\lim_{n\rightarrow \infty }\mathbb{P}%
_{n}\left\{ KS_{n}^{^{ldte}}>c_{\alpha ,n}^{^{ldte,\ast }}\right\} =\alpha .$

\item Under $H_{1}^{^{ldte}}$, $\lim_{n\rightarrow \infty }\mathbb{P}%
_{n}\left\{ KS_{n}^{^{ldte}}>c_{\alpha ,n}^{^{ldte,\ast }}\right\} =1.$

\item Under $H_{1,n}^{^{ldte}}:\Upsilon ^{^{ldte}}\left( y|\mathbf{X}\right)
=\frac{1}{\sqrt{n}}h^{^{ldte}}\left( y,\mathbf{X}\right) ~a.s.$ $\forall
y\in \mathcal{[-\infty ,}\tau \mathcal{]}$, if $h^{^{ldte}}\left( \cdot
,\cdot \right) $ is an integrable function and the set $h_{n}^{^{ldte}}%
\equiv \left\{ \left( y,\mathbf{x}\right) \in \mathcal{W}%
:n^{-1/2}h^{^{ldte}}\left( y,\mathbf{x}\right) \not=0\right\} $ has positive
Lebesgue measure, then $\lim_{n\rightarrow \infty }\mathbb{P}_{n}\left\{
KS_{n}^{^{ldte}}>c_{\alpha ,n}^{^{ldte,\ast }}\right\} >\alpha .$
\end{enumerate}
\end{theorem}

The results of Theorem \ref{tests_ldte} are related to \cite{Abadie2002}. In
the absence of censoring, \cite{Abadie2002} proposes a test for the
unconditional analogue of $H_{0}^{^{ldte}}$. Of course, by taking $w\left( 
\mathbf{X},\mathbf{x}\right) =1~a.s.$, we are back to \cite{Abadie2002}
proposal. Thus, one may interpret Theorem \ref{tests_ldte} as extensions of 
\cite{Abadie2002} in two different dimensions: it allows for covariates and
for randomly censored outcomes. We are not aware of other proposals that can
accommodate either these features.

\section{Monte Carlo simulations\label{MCsimul}}

In this section, we conduct a small-scale Monte Carlo exercise in order to
study the finite sample properties of our test statistics for the null
hypotheses (\ref{cste}), (\ref{h0-cate}) and (\ref{cctae}). The $\left\{
V_{i}\right\} _{i=}^{n}$ used in the bootstrap implementations are
independently generated as $V$ with $\mathbb{P}\left( V=1-\kappa \right)
=\kappa /\sqrt{5}$ and $\mathbb{P}\left( V=\kappa \right) =1-\kappa /\sqrt{5}
$, where $\kappa =\left( \sqrt{5}+1\right) /2$, as proposed by \cite%
{Mammen1993}. The bootstrap critical values are approximated by Monte Carlo
using $1,000$ replications and the simulations are based on $10,000$ Monte
Carlo experiments. We report rejection probabilities at the $5\%$
significance level. Results for 10\% and 1\% significance levels are similar
and available upon request.

We consider the following three designs:%
\begin{align*}
\left( i\right) .~~Y\left( 0\right) & =1+X+\varepsilon \left( 0\right)
,~Y\left( 1\right) =1+X+\varepsilon \left( 1\right) , \\
~~C\left( 0\right) & =C\left( 1\right) \sim a_{1}+b_{1}\times
Exponential\left( 1\right) ; \\
\left( ii\right) .~~Y\left( 0\right) & =1+X+e\left( 0\right) ,~Y\left(
1\right) =2+X+e\left( 1\right) , \\
~~C\left( 0\right) & =C\left( 1\right) \sim a_{2}+b_{2}\times
Exponential\left( 1\right) ; \\
\left( iii\right) .~~Y\left( 0\right) & =1+X+e\left( 1\right) ,~Y\left(
1\right) =1+3X+e\left( 1\right) , \\
~~C\left( 0\right) & =C\left( 1\right) \sim a_{3}+b_{3}\times
Exponential\left( 1\right) ;
\end{align*}%
where $X$ is distributed as $U\left[ 0,1\right] $, independently of $e\left(
0\right) ,e\left( 1\right) ,C\left( 0\right) $ and $C\left( 1\right) $, $%
\varepsilon \left( 0\right) $ and $\varepsilon \left( 1\right) $ are
independent standard normal random variables and the parameters $a$ and $b$
are chosen such that the percentage of censoring is equal to 0, 10 or 30
percent in the whole sample. In all designs, $\mathbb{P}\left( T=1|X\right)
=\exp \left( -0.5X\right) /\left( 1+\exp \left( -0.5X\right) \right) .$ When
testing (\ref{cste}) and (\ref{h0-cate}), Design $\left( i\right) $ falls
under the null, whereas Designs $\left( ii\right) -\left( iii\right) $ fall
under the alternative. When testing (\ref{cctae}), Designs $\left( i\right)
-\left( ii\right) $ fall under the null and Design $\left( iii\right) $
falls under the alternative. We set $\bar{\tau}=\tau $ when testing (\ref%
{h0-cate}) and (\ref{cctae}), implying that we do not rely on any truncation.

We report the proportion of rejections for sample sizes $n=100$, $300$ and $%
500$. We estimate $p\left( \cdot \right) $ using the SLE: with $n=100$ we
use $1,X$, with $n=300$ we use $1,X,X^{2}$ and with $n=500$ we use $%
1,X,X^{2},X^{3}$ as power functions in the estimation procedure. The
proportion of rejections for our tests are presented in Table \ref{tab:CDTE0}%
. $KS_{n}$ and $CvM_{n}$ stand for the $KS$ and $CvM$ test statistics for
the null of zero conditional distribution treatment effect. $%
KS_{n}^{^{cate}} $ and $CvM_{n}^{^{cate}}$ are the analogous test statistics
for the null of zero conditional average treatment effect and $%
KS_{n}^{^{\hom }}$ and $CvM_{n}^{^{\hom }}$for the null of homogeneous
average treatment effect across covariate values.

\begin{table}[]
\caption{Empirical rejection probabilities, in percentage points.}\centering%
\begin{adjustbox}{width=0.75\textwidth}
\label{tab:CDTE0}
\begin{threeparttable}
    \begin{tabular}{rcrrrrrrr} \hline
    \toprule
    \multicolumn{1}{c}{$DGP$} & \multicolumn{1}{c}{Censoring} & \multicolumn{1}{c}{$n$} & \multicolumn{1}{c}{$KS_{n}$} & \multicolumn{1}{c}{$CvM_{n}$} & \multicolumn{1}{c}{$KS_{n}^{cate}$} & \multicolumn{1}{c}{$CvM_{n}^{cate}$} & \multicolumn{1}{c}{$KS_{n}^{hom}$} & \multicolumn{1}{c}{$CvM_{n}^{hom}$} \\
    \midrule
   $(i)$     & 0   & 100   & 5.38 & 5.27 & 5.33 & 4.97 & 5.42 & 4.91 \\
    $(i)$     & 10  & 100   & 5.32 & 5.07 & 4.80 & 4.74 & 5.15 & 4.82 \\
    $(i)$     & 30  & 100   & 3.79 & 5.46 & 4.07 & 4.35 & 3.72 & 3.92 \\
   $(ii)$    & 0   & 100   & 97.52 & 98.50 & 99.04 & 98.93 & 5.85 & 5.10 \\
   $(ii)$    & 10  & 100   & 97.27 & 98.43 & 95.24 & 94.40 & 4.70 & 4.54 \\
   $(ii)$    & 30  & 100   & 76.28 & 95.86 & 52.74 & 52.98 & 4.19 & 4.14 \\
  $(iii)$   & 0   & 100   & 94.78 & 89.51 & 97.18 & 89.54 & 27.72 & 48.22 \\
 $(iii)$   & 10  & 100   & 92.33 & 86.65 & 91.58 & 80.91 & 16.22 & 27.32 \\
 $(iii)$   & 30  & 100   & 73.57 & 78.96 & 61.08 & 52.45 & 7.11 & 9.84 \\ \hline
  $(i)$     & 0   & 300   & 5.33 & 5.00 & 5.45 & 5.34 & 5.44 & 5.54 \\
 $(i)$     & 10  & 300   & 5.31 & 5.10 & 4.94 & 4.59 & 4.73 & 4.32 \\
 $(i)$     & 30  & 300   & 4.34 & 5.48 & 3.99 & 4.44 & 3.79 & 4.28 \\
  $(ii)$    & 0   & 300   & 100 & 100 & 100 & 100 & 5.16   & 4.83 \\
  $(ii)$    & 10  & 300   & 100 & 100 & 100 & 100 & 4.74   & 4.68 \\
  $(ii)$    & 30  & 300   & 99.51 & 100 & 92.81 & 92.53 & 4.17 & 4.39 \\
 $(iii)$   & 0   & 300   & 100 & 100 & 100 & 100 & 94.27  & 99.42 \\
 $(iii)$   & 10  & 300   & 100 & 100 & 100 & 100 & 66.50  & 84.70 \\
 $(iii)$   & 30  & 300   & 99.81 & 99.95 & 97.67 & 95.02 & 22.42 & 33.66 \\ \hline
 $(i)$     & 0   & 500   & 5.04 & 5.32 & 5.31 & 5.20 & 5.66 & 5.53 \\
 $(i)$     & 10  & 500   & 5.21 & 4.93 & 5.17 & 4.95 & 5.02 & 4.61 \\
 $(i)$     & 30  & 500   & 4.62 & 5.38 & 4.14 & 4.35 & 4.45 & 4.34 \\
  $(ii)$    & 0   & 500   & 100 & 100 & 100 & 100 & 5.61   & 5.13 \\
  $(ii)$    & 10  & 500   & 100 & 100 & 100 & 100 & 5.05   & 4.72 \\
  $(ii)$    & 30  & 500   & 99.96 & 100 & 98.77 & 98.53 & 4.42 & 4.79 \\
 $(iii)$   & 0   & 500   & 100 & 100 & 100 & 100 & 100 & 100 \\
 $(iii)$   & 10  & 500   &  100     &    100   & 100      & 99.99      & 91.41       &   97.76\\
 $(iii)$   & 30  & 500   &  99.99    &  100  & 99.78  & 99.22      & 39.03      & 53.51 \\
  \bottomrule \hline
\end{tabular}%

\end{threeparttable}
\end{adjustbox}
\end{table}

We observe that our tests exhibit good size accuracy even when $n=100$. When
the censoring level is 30\%, we have that the proposed tests have size below
their nominal levels, but as we increase the sample size, such size
distortions are minimized. With respect to power, our $KS$ and $CvM$ test
statistics reach satisfactory levels for $n=100$, the only exception being
when testing for homogeneous ATE with\ a censoring level of 30\%.
Nonetheless, as the sample size increases, all tests present satisfactory
power properties, regardless of the censoring level considered. As one
should expect, the power of all tests increases with sample size, and
decreases with the degree of censoring. Overall, these simulations show that
the proposed bootstrap tests exhibit excellent finite sample properties.

\section{Illinois Reemployment Bonus Experiment\label{Illinois}}

In this section, we demonstrate that our proposed tests can be useful in
practice. We analyze data from the Illinois Reemployment Bonus Experiments,
which is freely available at the W.E. Upjohn Institute for Employment
Research.

From mid-1984 to mid-1985, the Illinois Department of Employment Security
conducted a social experiment to test the effectiveness of bonus offers in
reducing the duration of insured unemployment. \ At the beginning of each
claim, the experiment randomly divided newly unemployed people into three
groups:

\begin{enumerate}
\item Job Search Incentive Group (JSI). The members of this group were told
that they would qualify for a cash bonus of \$500, which was about four
times the average weekly unemployment insurance benefits, if they found a
full-time job within eleven weeks of benefits and if they held that job for
at least four months. 4816 claimants were assigned to this group.

\item Hiring Incentive Group (HI). The members of this group were told that
their employer would qualify for a cash bonus of \$500 if the claimant found
a full-time job within eleven weeks of benefits and if they held that job
for at least four months. 3963 claimants were assigned to this group.

\item Control Group. All claimants not assigned to the other groups. These
members did not know that the experiment was taking place. 3952 individuals
were assigned to this group.
\end{enumerate}

An important aspect of the Illinois Reemployment Bonus Experiment is that
participation was not mandatory. Once claimants were assigned to the
treatment groups, they were asked if they would like to participate in the
demonstration or not. For those selected to the Job Search Incentive group,
84\% agreed to participate, whereas just 65\% of the Hiring Incentive group
agreed to participate.

Several studies including \cite{Woodbury1987}, \cite{Meyer1996}, and \cite%
{Bijwaard2005} have analyzed the impact of the reemployment bonus on the
unemployment duration measured by the number of weeks receiving unemployment
insurance. Spells which reached the maximum amount of benefits or the state
maximum number of weeks, 26, are censored, leading to censoring proportions
of 38, 41 and 42 percent for the JSI, HI and the control group,
respectively. Apart from the duration data, some information about
claimants' background characteristics is also available: age, gender (Male
=1), ethnicity (White =1), pre-unemployment earnings and the weekly
unemployment insurance benefits amount. For a complete description of the
experiment and the available dataset, see \cite{Woodbury1987}.

Our goal in this application is to assess the effect of reemployment bonuses
on unemployment duration. Given the differences between JSI and HI, we
analyze these two treatments separately. That is, we consider two
sub-samples: one with individuals who are in JSI or in the control group,
and one with individuals who are in HI or in the control group.\footnote{%
 Given that the allocation to each treatment arm is random, i.e.,
the data come from a randomized control trial, there is no loss of
generality of restricting our attention to these two sub-samples. In cases
where one has more than two treatment groups and selection into these groups
is not completely random, we foresee that one would be able to combine the
tools developed in this paper with those in \cite{Cattaneo2010} and \cite%
{Ao2019}, under the unconfounded setup. When treatment is endogenous,
though, it is not yet clear how one can extend the analysis to accomodate
multiple instruments; see \cite{Mogstad2019} for some recent results in this
direction. We leave a detailed discussion of these extensions to future
research.} Furthermore, we consider two types of analysis. First, we
consider an intention to treat (ITT) analysis, where $T=1$ if an individual
is offered to participate in the demonstration, and $T=0$ if an individual
was in the control group. In this case, we completely ignored the
non-compliance with treatment allocations. Second, in an attempt to
disentangle the effects of being offered and actually receiving treatment,
we consider a local treatment effect analysis, using the random assignment
as an instrumental variable.

Table \ref{tab:illinois} reports the results of all our proposed tests,
based on 10,000 bootstrap replications. We consider the nulls of $(a)$ zero
conditional (local) DTE, $(b)$ zero conditional (local) ATE, and $(c)$
homogeneous (local) ATE across covariate values. The conditioning vector
considered consists of all available claimants characteristics described
above.

To implement all tests, we estimate the propensity score $p_{0}\left( \cdot
\right) $ and the instrument propensity score $q_{0}\left( \cdot \right) $
using the SLE where all covariates enter the model linearly. Given that the
data comes from an experimental design, consistency of the propensity score
models is guaranteed.

Let us start interpreting the results for the JSI sample. For both the ITT
and LTE setup, we reject the null of zero conditional (local) DTE at the 5\%
level using either the Kolmogorov-Smirnov or the Cram\'{e}r-von Mises test
statistic. Such evidence suggests that offering a reemployment bonus to
job-searchers has affected the distribution of unemployment duration.

\begin{table}[]
\caption{Bootstrap p-values for different tests for treatment effect
heterogeneity based on the Illinois bonus experiment. }\centering
\begin{adjustbox}{width=0.85\textwidth}
\label{tab:illinois}
\begin{threeparttable}

  \begin{tabular}{lcccc}
    \toprule \hline
    \multicolumn{5}{c}{Intention to Treat } \\ \hline
    \midrule
          & \multicolumn{2}{c}{Job Search Incentive Group} & \multicolumn{2}{c}{Hiring Incentive Group} \\ \hline
    Null Hypothesis /  Test type& $KS$    & $CvM$   & $KS$    & $CvM$ \\ \hline
    Zero Conditional DTE & 0.0001     & 0.0122 & 0.0618 & 0.1516 \\
    Zero Conditional ATE & 0.0335 & 0.0686 & 0.1533 & 0.2167 \\
    Homogeneous Conditional ATE & 0.0319 & 0.0992 & 0.5568 & 0.6505 \\ \hline \hline
    \multicolumn{5}{c}{Local Treatment Effects - Compliers} \\ \hline
          & \multicolumn{2}{c}{Job Search Incentive Group} & \multicolumn{2}{c}{Hiring Incentive Group} \\ \hline
    Null Hypothesis / Test type & $KS$    & $CvM$   & $KS$    & $CvM$ \\ \hline
    Zero Conditional Local DTE & 0.0001 & 0.0105 & 0.0598 & 0.1559 \\
    Zero Conditional Local ATE & 0.0386 & 0.0766 & 0.1589 & 0.2191 \\
    Homogeneous Conditional Local ATE & 0.5910 & 0.5851 & 0.9949 & 0.9725 \\ \hline
    \bottomrule
    \end{tabular}%

\end{threeparttable}
\end{adjustbox}
\end{table}

To shed some light on which part of the distribution is affected, we test
the null of zero conditional (restricted) ATE as in (\ref{h0-cate}). In the
LTE setup, we consider the analogous null of zero conditional local
(restricted) ATE 
\begin{equation*}
H_{0}^{^{clate}}:\Upsilon _{\bar{\tau}}^{^{clate}}\left( \mathbf{X}\right)
=0~a.s.,
\end{equation*}%
where%
\begin{eqnarray*}
\Upsilon _{\bar{\tau}}^{^{clate}}\left( \mathbf{X}\right) &=&\mathbb{E}\left[
Y\left( 1\right) 1\left\{ Y\left( 1\right) \leq \bar{\tau}\right\} |\mathbf{X%
},pop=comp\right] \\
&&-\mathbb{E}\left[ Y\left( 0\right) 1\left\{ Y\left( 0\right) \leq \bar{\tau%
}\right\} |\mathbf{X},pop=comp\right] .
\end{eqnarray*}%
For details about how one can construct tests for $H_{0}^{^{clate}}$, see
Section S.1.1 of the Supplementary Appendix.

We set $\bar{\tau}=26$, so all the available data is used. From Table \ref%
{tab:illinois}, we have that $H_{0}^{^{cate}}$ and $H_{0}^{^{clate}}$ are
both rejected at the 5\% level when using the $KS$ test and at the 10\%
level when using the $CvM$ test. Such evidence suggests that the
reemployment bonus affected the average unemployment duration. However, we
note that \cite{SantAnna2016} unconditional (restricted) ATE and LATE
estimators are $-0.2221$ and $1.9745$, respectively, and both are not
statistically significant at the 10\% level. Thus, a researcher who relied
only on \textquotedblleft traditional\textquotedblright\ unconditional tests
of a zero average effect would have missed the presence of treatment effects
in this program.

When conditional (local) ATE is heterogeneous, it may be harder to identify
the subpopulation of individuals that the treatment effect is non-zero.
However, if the conditional ATE is homogeneous, the task is trivial. With
this in mind, we test the null of homogenous conditional (restricted) ATE as
in (\ref{cctae}). In the LTE setup, we consider the analogous null of
homogenous conditional local (restricted) ATE,%
\begin{equation*}
H_{0}^{^{l\hom }}:\exists ~\Upsilon _{\bar{\tau}}^{l}\in \mathbb{R}:\Upsilon
_{\bar{\tau}}^{^{clate}}\left( \mathbf{X}\right) =\Upsilon _{\bar{\tau}%
}^{l}~a.s..
\end{equation*}%
For details about how one can construct tests for $H_{0}^{^{l\hom }}$, see
Section S.1.2 of the Supplementary Appendix.

As before, we set $\bar{\tau}=26$. For the ITT setup, the null of
homogeneous conditional ATE is rejected at the 5\% level when using the $KS$
test, and at the 10\% level when using the $CvM$ test. When the endogeneity
of the selection into treatment is taken into account, we fail to reject the
null of homogenous conditional local (restricted) ATE at usual confidence
levels. From these results, one concludes that the average treatment effect
of being offered versus not being offered into the bonus experiment is
heterogeneous. On the other hand, once we restrict our attention to the
complier subpopulation, we fail to find enough evidence against the null of
homogeneous ATE of actually participating in the JSI program versus not
participating.

Next, we analyze the results for the HI sub-sample. Interesting enough, at
the 5\% level, we fail to reject each considered null hypothesis regardless
of the test statistic used. This finding suggests that offering a
reemployment bonus to the employer does not affect the time unemployed
individuals take to find a job at all.

Overall, the results of our proposed tests suggest that offering an
unemployment bonus to the job searcher was effective in changing the length
of the unemployment spell. On the other hand, offering the bonus to the
employer rather than to the job-searcher seems to be ineffective in changing
the unemployment duration.

\section{Conclusion and Suggestions for further Work}

In this article, we proposed a variety of nonparametric tests for treatment
effect heterogeneity that can accommodate randomly censored outcomes and
endogenous treatment allocations. We derived the asymptotic properties of
the proposed tests, and have proved that critical values can be easily
computed via a relatively simple multiplier bootstrap procedure.
Furthermore, in contrast to other proposals, we proved that our tests are
able to detect local alternatives converging to the null at the parametric
rate. Our Monte Carlo simulations show that our proposed tests have good
finite sample properties. Finally, our empirical application concerning the
effect of unemployment bonus on unemployment duration showed the feasibility
and appeal of our tests in relevant scenarios. Given the desirable features
of our tests and the importance of treatment effect heterogeneity in
assessing external validity, we argue that the tests proposed in this
article are important additions to the applied researcher's toolkit.

For concreteness and compatibility of the procedures for both duration and
non-duration outcomes, we framed the article within the context of
time-invariant treatment allocation, i.e. when treatment allocation happens
at beginning of the duration spell. Nonetheless, when treatment allocation
is dynamic, the results of this article still apply to testing for treatment
effect heterogeneity between those individuals treated at time $t$ and those
not yet treated at time $t,$ as in \cite{Sianesi2004}. Once the treatment
and control groups are defined, the implementation of our proposed tests are
exactly the same as described in the text.

In the rest of this section, we discuss extensions of our proposed
methodology to other situations of practical interest. For example, we note
that researchers may be interested in tests for conditional stochastic
dominance, or more generally, tests based on conditional moment
inequalities, see e.g. \cite{Andrews2013, Andrews2017}, \cite{Delgado2013}, 
\cite{Chang2015} and \cite{Hsu2013}. As discussed in Section 5.1 of \cite%
{Andrews2013}, pointwise asymptotics results do not provide good
approximations to the finite-sample properties of test statistics in
conditional moment inequality models. Thus, in order to consider tests based
on conditional moment inequalities under random censoring, one should first
establish a uniform in DGP Bahadur expansion of a suitable two-step
Kaplan-Meier integrals; for setting without censoring, see e.g. Lemma 3.1
and Lemma 3.2 in \cite{Hsu2013}. However, as noted in Remark \ref{one-sided}%
, establishing such uniform Bahadur representation for Kaplan-Meier
integrals is very technically challenging because traditional empirical
process tools such as those in chapter 2.8 of \cite{VanderVaart1996} and in
chapter 10 of \cite{Dudley2014} are not directly applicable to Kaplan-Meier
processes. Indeed, to the best of our knowledge, no such result is yet
available in the literature, even for the classical univariate Kaplan-Meier
estimator for the CDF, let alone more general classes of functions. In light
of this observation, we argue that a formal treatment of tests based on
conditional moment inequalities under random censoring is beyond the scope
of this article and is left for future research.

Another interesting extension one may whish to pursue is constructing tests
for treatment effect heterogeneity with randomly right-censored outcomes
that impose less stringent conditions on the censoring mechanism than those
allowed under Assumption \ref{censoring_identification}. For instance, one
may wish to assume that, for $t\in \left\{ 0,1\right\} $, $\left( Y\left(
0\right) ,Y\left( 1\right) \right) $ $%
%TCIMACRO{\TeXButton{indep}{\independent}}%
%BeginExpansion
\independent%
%EndExpansion
\left( C\left( 0\right) ,C\left( 1\right) \right) |T,\mathbf{X}$, and
propose tests based on conditional Kaplan-Meier estimators using procedures
similar to \cite{Akritas1994}, \cite{Gonzalez-Manteiga1994} and \cite%
{Lopez2011}. Although such an assumption is less restrictive than Assumption %
\ref{censoring_identification}, constructing tests for treatment effect
heterogeneity based on conditional Kaplan-Meier estimators would rely on an
additional set of assumptions. For instance, it would $\left( a\right) $
require additional stronger smoothness and differentiability assumptions; $%
\left( b\right) $ rule out empirically relevant situations with discrete
duration data as in our empirical application; $\left( c\right) $ involve
choosing additional tuning parameters to estimate the conditional CDF; $%
\left( d\right) $ and perhaps even more importantly, the statistical
analysis would rely on truncation arguments, which in turn would exclude
from the analysis some important classes of functions such as (conditional)
average treatment effects. At the cost of restricting how covariates may
affect the probability of being censored, the tests proposed in this article
bypass all these additional challenges. Given that these two procedures rely
on different non-nested assumptions, and involves a very different type of
statistical analysis, it is hard to generally rank them. We leave such task
for future research.

\pagebreak \onehalfspacing%\small
\begin{appendices}

\setcounter{equation}{0} 
\titleformat{\section}{\center \Large
\bfseries}{Appendix \thesection: }{0.1cm}{} 
\titleformat{\subsection}{\large
\bfseries}{\thesubsection}{1em}{}
\numberwithin{lemma}{section}
\numberwithin{theorem}{section}

\section{Technical Assumptions \label{App-Assumption}}

{We first present the technical Assumptions needed for our main results. Let 
$\mathcal{C}_{b}\left( \mathbb{R}^{k}\right) $ be the space of all bounded,
continuous, complex-valued functions on $\mathbb{R}^{k}$. }

\begin{assumption}
{\ \label{w-family}The class of functions $\mathcal{F}=\left\{ w\left( 
\mathbf{X},\mathbf{x}\right) :\mathbf{x}\in \Pi \subset \left[ -\infty
,\infty \right] ^{k}\right\} $ satisfy one of the following conditions: }

\begin{enumerate}
\item[$\left( i\right) $] {\ $\mathcal{F\subset C}_{b}\left( \mathbb{R}%
^{k}\right) $ is a vector lattice that contains the constant functions and
separates points of $\mathbb{R}^{k}.$ }

\item[$\left( ii\right) $] {\ $\mathcal{F\subset C}_{b}\left( \mathbb{R}%
^{k}\right) $ is an algebra that contains the constant functions and
separates points of $\mathbb{R}^{k}.$ }

\item[$\left( iii\right) $] {\ $\mathcal{F}=\left\{ w\left( \mathbf{x}%
^{\prime }\mathbf{X}\right) :\mathbf{x}\in \Pi \subset \left[ -\infty
,\infty \right] ^{k}\right\} $ and $w$ is an analytic function that is
non-polynomial, where }${\ \Pi }${\ \ is a compact set of $\mathbb{R}^{k}$
containing the origin. }

\item[$\left( iv\right) $] {\ $\mathcal{F=}\left\{ 1\left( \mathbf{X}\in
B_{x}\right) :\mathbf{x}\in \Pi \subset \left[ -\infty ,\infty \right]
^{k}\right\} $ and $\left\{ B_{x}\right\} _{x\in \mathcal{X}_{_{X}}}$ is a
separating class of Borel sets of $\mathbb{R}^{k}$. }
\end{enumerate}
\end{assumption}

{\ Assumption \ref{w-family} states the conditions on $w$ such that Lemma %
\ref{integrated} holds. These conditions are exactly the same as those in 
\cite{Escanciano2006a} Lemma 1. }

\begin{assumption}
{\ \label{support_pscore_series} }

\begin{enumerate}
\item[(i)] {\ The support }$\mathcal{X}${$_{_{_{X}}}$ of the $k$-dimensional
covariate }${\mathbf{X}}${\ \ is a Cartesian product of compact intervals, }$%
\mathcal{X}$$_{_{_{X}}}${$=\prod_{j=1}^{k}\left[ x_{lj},x_{uj}\right] ;$ }

\item[(ii)] {\ The density of }${\mathbf{X}}${\ \ is bounded, and bounded
away from 0 on }$\mathcal{X}$$_{_{_{X}}}${\ }
\end{enumerate}
\end{assumption}

\begin{assumption}
{\ \label{diff-dte}For $t\in \left\{ 0,1\right\} $, $F_{Y\left( t\right)
|X}\left( y|\mathbf{X}=\mathbf{x}\right) $ is $m$-times continuously
differentiable in $x$, for all }$\left( y,\mathbf{x}\right) \in \mathcal{X}${%
$_{_{_{Y}}}$}$\times \mathcal{X}$$_{_{_{X}}}${, $m\geq k$. }
\end{assumption}

\begin{assumption}
{\ \label{derivatives_pscore}For all $\mathbf{x}\in $}$\mathbb{X}${{$_{_{X}}$%
}, the propensity score $p_{0}\left( \mathbf{x}\right) $ is continuously
differentiable of order $s\geq 13k$, where $k$ is the dimension of $X$. }
\end{assumption}

\begin{assumption}
{\ \label{series estimator} The series logit estimator of $p_{0}\left( 
\mathbf{x}\right) $ uses a power series with $L=a\cdot N^{v}$ for some $a>0$
and $1/\left( s/k-2\right) <v<1/11$. }
\end{assumption}

\begin{assumption}
{\ \label{moments_dte}For $t\in \left\{ 0,1\right\} $, assume that, for all $%
\left( y,\mathbf{x}\right) \in \mathcal{W,}$ 
\begin{eqnarray*}
\mathbb{E}\left[ \left( 1\left\{ Q\left( t\right) \leq y\right\} 1\left\{ 
\mathbf{X}\leq \mathbf{x}\right\} \gamma _{t,0}\left( Q\right) \delta
_{t}\right) ^{2}\right] &<&\infty , \\
\mathbb{E}\left[ 1\left\{ Q\left( t\right) \leq y\right\} 1\left\{ \mathbf{X}%
\leq \mathbf{x}\right\} C_{t}^{1/2}\left( Y\right) \right] &<&\infty ,
\end{eqnarray*}%
where $\gamma _{t,0}$ is defined as in (\ref{gam0}), 
\begin{equation*}
C_{t}\left( w\right) =\int_{-\infty }^{w-}\frac{G_{t}(dy)}{\left[
1-H_{t}\left( y\right) \right] \left[ 1-G_{t}\left( y\right) \right] },
\end{equation*}%
and $G_{t}\left( w\right) =\mathbb{P}\left( C\leq w,T=t\right) $. }
\end{assumption}

\begin{assumption}
{\ \label{diff-ate}For $t\in \left\{ 0,1\right\} $, $E\left( Y\left(
t\right) |\mathbf{X}=\mathbf{x}\right) $ is m-times continuously
differentiable in $x$, for all $\mathbf{x}\in \mathcal{X}{_{_{_{X}}}},$ $%
m\geq k.$ }
\end{assumption}

\begin{assumption}
{\ \label{moments_ate}For $t\in \left\{ 0,1\right\} $, assume that, for all $%
\left( y,\mathbf{x}\right) \in \mathcal{W,}$ 
\begin{equation*}
\mathbb{E}\left[ \left( Q\left( t\right) 1\left\{ \mathbf{X}\leq \mathbf{x}%
\right\} \gamma _{t,0}\left( Q\right) \delta _{t}\right) ^{2}\right] <\infty
,
\end{equation*}%
\begin{equation}
\mathbb{E}\left[ Q\left( t\right) 1\left\{ \mathbf{X}\leq \mathbf{x}\right\}
C_{t}^{1/2}\left( Y\right) \right] <\infty .
\end{equation}%
}
\end{assumption}

{\ Similar assumptions have adopted by \cite{Hirano2003}, \cite{Crump2008},\ 
\cite{Donald2013}, among others. Assumptions \ref{support_pscore_series}, %
\ref{diff-dte} and \ref{diff-ate} restrict the distribution of }$\mathbf{X}${%
\ and $Y\left( t\right) $ and requires that all covariates are continuous.
Nonetheless, at the expense of additional notation, we can deal with the
case where }$\mathbf{X}$ {has both continuous and discrete components by
means of sample splitting based on the discrete covariates. In order to
avoid cumbersome notation, we abstract from this point in the rest of the
paper. Assumption \ref{derivatives_pscore} requires sufficient smoothness of
the propensity score, whereas Assumption \ref{series estimator} restrict the
rate at which additional terms are added to the series approximation of $%
p\left( \cdot\right) $, depending on the dimension of }$\mathbf{X}${\ \ and
the number of derivatives of $p\left(\cdot\right) $. The restriction on the
derivatives in Assumption \ref{derivatives_pscore} guarantees the existence
of a $v$ that satisfy the conditions in Assumption \ref{series estimator}.
Assumptions \ref{moments_dte} and \ref{moments_ate} are standard with
censored data; they guarantee that the variance of the Kaplan-Meier integral
related to the DTE and ATE is finite, and their bias are $o\left(
n^{-1/2}\right) $. See \cite{Stute1996a} and \cite{Chen1997} for a detailed
discussion. }

\section{Details About the Tests of Section 4\label{App-B}}

In Section \ref{Extensions}, we discuss extensions of our basic setup. More
formally, we proposed tests for the null of zero conditional average
treatment effect and for the null of constant average treatment effect
across subpopulations. Furthermore, we showed how one can modify the
aforementioned tests to accommodate endogenous treatment allocation. In this
Appendix, we provide more details on how to construct the test statistics.

As discussed in Section \ \ref{ate-sec}, to test the null of zero
conditional average treatment effect 
\begin{equation*}
H_{0}^{^{cate}}:\Upsilon _{\bar{\tau}}^{^{cate}}\left( \mathbf{X}\right) =0%
\text{ }a.s.,
\end{equation*}%
one can use the $KS$-type test%
\begin{equation*}
KS_{\bar{\tau},n}^{^{cate}}=\sup_{\mathbf{x}\in \mathcal{X}_{X}}\left\vert 
\sqrt{n}\hat{I}_{\bar{\tau},n}^{^{cate}}(\mathbf{x})\right\vert ,
\end{equation*}%
where 
\begin{equation}
\hat{I}_{\bar{\tau},n}^{^{cate}}\left( y,\mathbf{x}\right) =\hat{I}_{\bar{%
\tau},n}^{^{1,cate}}\left( x\right) -\hat{I}_{\bar{\tau},n}^{^{0}}\left( 
\mathbf{x}\right) ,  \label{Icate}
\end{equation}%
with 
\begin{equation*}
\hat{I}_{\bar{\tau},n}^{^{t,cate}}\left( \mathbf{x}\right)
=\sum_{i=1}^{n_{t}}W_{in_{t}}\frac{Q_{i:n_{t}}1\left\{ Q_{i:n_{t}}\leq \bar{%
\tau}\right\} 1\left( \mathbf{X}_{\left[ i:n_{t}\right] }\leq \mathbf{x}%
\right) }{\mathbb{\hat{P}}_{n}\left( T=t|\mathbf{X}_{\left[ i:n_{t}\right]
}\right) },~t\in \left\{ 0,1\right\} ,
\end{equation*}%
$\mathbb{\hat{P}}_{n}\left( T=t|\mathbf{X}\right) $ the Series Logit
Estimator for $\mathbb{P}\left( T=t|\mathbf{X}\right) ,$ $t\in \left\{
0,1\right\} $.

To test the null of homogeneous average treatment effect, 
\begin{equation*}
H_{0}^{^{\hom }}:\exists ~\Upsilon _{\bar{\tau}}\in \mathbb{R}:\Upsilon _{%
\bar{\tau}}^{^{cate}}\left( \mathbf{X}\right) =\Upsilon _{\bar{\tau}}\text{ }%
a.s..
\end{equation*}%
we propose test statistic 
\begin{equation*}
KS_{\bar{\tau},n}^{^{\hom }}=\sup_{\mathbf{x}\in \mathcal{X}_{X}}\left\vert 
\sqrt{n}\hat{I}_{\bar{\tau},n}^{^{\hom }}(\mathbf{x})\right\vert ,
\end{equation*}%
where 
\begin{equation}
\hat{I}_{\bar{\tau},n}^{^{\hom }}\left( \mathbf{x}\right) =\hat{I}_{\bar{\tau%
},n}^{^{1,\hom }}\left( \mathbf{x}\right) -\hat{I}_{\bar{\tau},n}^{^{0,\hom
}}\left( \mathbf{x}\right) ,  \label{Ihom}
\end{equation}%
with, for $t\in \left\{ 0,1\right\} $, 
\begin{equation*}
\hat{I}_{\bar{\tau},n}^{^{t,\hom }}\left( \mathbf{x}\right) =\frac{n_{t}}{n}%
\sum_{i=1}^{n_{t}}W_{in_{t}}\left( \frac{Q_{i:n_{t}}1\left\{ Q_{i:n_{t}}\leq 
\bar{\tau}\right\} }{\mathbb{\hat{P}}_{n}\left( T=t|\mathbf{X}_{\left[
i:n_{t}\right] }\right) }-\left( 2t-1\right) \hat{I}_{\bar{\tau}%
,n}^{^{ate}}\right) 1\left( \mathbf{X}_{\left[ i:n_{t}\right] }\leq \mathbf{x%
}\right) ,
\end{equation*}%
$\hat{p}_{n}\left( \cdot \right) $ is the SLE for $p_{0}\left( \cdot \right) 
$, and 
\begin{equation*}
\hat{I}_{\bar{\tau},n}^{^{ate}}=\frac{n_{1}}{n}\sum_{i=1}^{n_{1}}W_{in_{1}}%
\frac{Q_{i:n_{1}}1\left\{ Q_{i:n_{1}}\leq \bar{\tau}\right\} }{\hat{p}%
_{n}\left( \mathbf{X}_{\left[ i:n_{1}\right] }\right) }-\frac{n_{0}}{n}%
\sum_{j=1}^{n_{0}}W_{jn_{0}}\frac{Q_{j:n_{0}}1\left\{ Q_{j:n_{0}}\leq \bar{%
\tau}\right\} }{1-\hat{p}_{n}\left( \mathbf{X}_{\left[ j:n_{0}\right]
}\right) }.
\end{equation*}

Finally, to test the null of zero local conditional distribution treatment
effect,%
\begin{equation*}
H_{0}^{^{ldte}}:\Upsilon ^{^{ldte}}\left( y|\mathbf{X}\right) =0\text{ }a.s.%
\text{ }\forall y\mathbf{\in }[\mathcal{-\infty ,}\tau ],
\end{equation*}%
we propose the $KS$ type test statistic%
\begin{equation*}
KS_{n}^{^{ldte}}=\sqrt{n}\sup_{\left( y,\mathbf{x}\right) \in \mathcal{W}%
}\left\vert \hat{I}_{n}^{^{ldte}}(y,\mathbf{x})\right\vert ,
\end{equation*}%
where%
\begin{equation}
\hat{I}_{n}^{^{ldte}}(y,\mathbf{x})=\hat{I}_{n}^{^{1,ldte}}(y,\mathbf{x})-%
\hat{I}_{n}^{^{0,ldte}}(y,\mathbf{x}),  \label{Ildte}
\end{equation}%
with%
\begin{eqnarray*}
\hat{I}_{n}^{^{1,ldte}}\left( y,\mathbf{x}\right) &=&\frac{n_{11}}{n}%
\sum_{i=1}^{n_{11}}W_{in_{11}}\frac{1\left\{ Q_{1:n_{11}}\leq y\right\}
1\left\{ \mathbf{X}_{\left[ i:n_{11}\right] }\leq \mathbf{x}\right\} }{\hat{q%
}_{n}\left( \mathbf{X}_{\left[ i:n_{11}\right] }\right) } \\
&&-\frac{n_{10}}{n}\sum_{i=1}^{n_{10}}W_{in_{10}}\frac{1\left\{
Q_{1:n_{10}}\leq y\right\} 1\left\{ \mathbf{X}_{\left[ i:n_{10}\right] }\leq 
\mathbf{x}\right\} }{1-\hat{q}_{n}\left( \mathbf{X}_{\left[ i:n_{10}\right]
}\right) }, \\
\hat{I}_{n}^{^{0,ldte}}\left( y,\mathbf{x}\right) &=&\frac{n_{00}}{n}%
\sum_{j=1}^{n_{00}}W_{jn_{00}}\frac{1\left\{ Q_{j:n_{00}}\leq y\right\}
1\left\{ \mathbf{X}_{\left[ j:n_{00}\right] }\leq \mathbf{x}\right\} }{1-%
\hat{q}_{n}\left( \mathbf{X}_{\left[ j:n_{00}\right] }\right) } \\
&&-\frac{n_{01}}{n}\sum_{j=1}^{n_{01}}W_{jn_{01}}\frac{1\left\{
Q_{j:n_{01}}\leq y\right\} 1\left\{ \mathbf{X}_{\left[ j:n_{01}\right] }\leq 
\mathbf{x}\right\} }{\hat{q}_{n}\left( \mathbf{X}_{\left[ j:n_{01}\right]
}\right) }
\end{eqnarray*}%
where $\hat{q}_{n}\left( \cdot \right) $ is the SLE for $q_{0}\left( \cdot
\right) $, $n_{tz}=\sum_{i=1}^{n}1\left\{ T=t\right\} 1\left\{ Z=z\right\} $%
, $t,z\in \left\{ 0,1\right\} $, and for $1\leq i\leq n_{tz},$ $%
Q_{1:n_{tz}}\leq $ $\cdots \leq Q_{n_{tz}:n_{tz}}$ are the ordered $Q$%
-values in the sub-sample with $\left\{ T=t,Z=z\right\} $, $\mathbf{X}_{%
\left[ i:n_{tz}\right] }$ and $\delta _{\left[ i:n_{tz}\right] }$ are the $%
\mathbf{X}$ and $\delta $ paired with $Q_{i:n_{tz}}$, and%
\begin{equation}
W_{in_{tz}}=\frac{\delta _{\lbrack i:n_{tz}]}}{n_{tz}-i+1}\prod_{j=1}^{i-1}%
\left[ \frac{n_{tz}-j}{n_{tz}-j+1}\right] ^{\delta _{\left[ j:n_{tz}\right]
}}  \label{w-lte}
\end{equation}%
is the Kaplan-Meier weights for the sub-sample with $\left\{ T=t,Z=z\right\} 
$.\bigskip

%{\footnotesize \onehalfspacing
%\bibliographystyle{jasa}
%\bibliography{JMP}
%}
\end{appendices}

{\footnotesize {\ 
\bibliographystyle{jasa}
\bibliography{JMP}
} }

\end{document}